\documentclass[superscriptaddress,twocolumn,
 amsmath,amssymb,
 aps,
 prx,
]{revtex4-2}

\usepackage{graphicx}
\usepackage{dcolumn}
\usepackage{bm}
\usepackage[detect-all]{siunitx}
\usepackage{color}
\usepackage[english]{babel}
\usepackage{ulem}

\usepackage{xcolor}
\definecolor{apsblue}{rgb}{0.176, 0.152, 0.57}
\usepackage[colorlinks=true, linkcolor=black, citecolor=apsblue, filecolor=black, urlcolor=apsblue]{hyperref}

\usepackage{xr}
\makeatletter
\newcommand*{\addFileDependency}[1]{
  \typeout{(#1)}
  \@addtofilelist{#1}
  \IfFileExists{#1}{}{\typeout{No file #1.}}
}
\makeatother
 
\newcommand*{\myexternaldocument}[1]{%
    \externaldocument{#1}%
    \addFileDependency{#1.tex}%
    \addFileDependency{#1.aux}%
}

\myexternaldocument{SM}
\begin{document}

\title{Stable and high quality electron beams from staged laser and plasma wakefield accelerators}

\author{F. M. Foerster}
\email{moritz.foerster@physik.uni-muenchen.de}
\affiliation{Ludwig--Maximilians--Universit{\"a}t M{\"u}nchen, Am Coulombwall 1, 85748 Garching, Germany}%

\author{A. D{\"o}pp}
\email{a.doepp@physik.uni-muenchen.de}
\affiliation{Ludwig--Maximilians--Universit{\"a}t M{\"u}nchen, Am Coulombwall 1, 85748 Garching, Germany}%
\affiliation{Max Planck Institut für Quantenoptik, Hans-Kopfermann-Strasse 1, 85748 Garching , Germany}%

\author{F. Haberstroh}
\affiliation{Ludwig--Maximilians--Universit{\"a}t M{\"u}nchen, Am Coulombwall 1, 85748 Garching, Germany}%

\author{K. v. Grafenstein}
\affiliation{Ludwig--Maximilians--Universit{\"a}t M{\"u}nchen, Am Coulombwall 1, 85748 Garching, Germany}%

\author{D. Campbell}
\affiliation{Ludwig--Maximilians--Universit{\"a}t M{\"u}nchen, Am Coulombwall 1, 85748 Garching, Germany}%
\affiliation{University of Strathclyde, 107 Rottenrow, Glasgow G4 0NG, United Kingdom}

\author{Y.-Y. Chang}
\affiliation{Helmholtz-Zentrum Dresden--Rossendorf, Bautzner Landstrasse 400, 01328 Dresden, Germany}%

\author{S. Corde}
\affiliation{Laboratoire d’Optique Appliquée, ENSTA Paris, CNRS, Ecole Polytechnique, Institut Polytechnique de Paris, 91762 Palaiseau, France}%

\author{J. P. Couperus Cabada{\u{g}}}
\affiliation{Helmholtz-Zentrum Dresden--Rossendorf, Bautzner Landstrasse 400, 01328 Dresden, Germany}%

\author{A. Debus}
\affiliation{Helmholtz-Zentrum Dresden--Rossendorf, Bautzner Landstrasse 400, 01328 Dresden, Germany}%

\author{M. F. Gilljohann}
\affiliation{Ludwig--Maximilians--Universit{\"a}t M{\"u}nchen, Am Coulombwall 1, 85748 Garching, Germany}%
\affiliation{Laboratoire d’Optique Appliquée, ENSTA Paris, CNRS, Ecole Polytechnique, Institut Polytechnique de Paris, 91762 Palaiseau, France}%

\author{A. F. Habib}
\affiliation{University of Strathclyde, 107 Rottenrow, Glasgow G4 0NG, United Kingdom}

\author{T. Heinemann}
\affiliation{University of Strathclyde, 107 Rottenrow, Glasgow G4 0NG, United Kingdom}
\affiliation{The Cockcroft Institute, Keckwick Lane, Warrington WA4 4AD, United Kingdom}%

\author{B. Hidding}
\affiliation{The Cockcroft Institute, Keckwick Lane, Warrington WA4 4AD, United Kingdom}%
\affiliation{University of Strathclyde, 107 Rottenrow, Glasgow G4 0NG, United Kingdom}

\author{A. Irman}
\affiliation{Helmholtz-Zentrum Dresden--Rossendorf, Bautzner Landstrasse 400, 01328 Dresden, Germany}%

\author{F. Irshad}
\affiliation{Ludwig--Maximilians--Universit{\"a}t M{\"u}nchen, Am Coulombwall 1, 85748 Garching, Germany}%

\author{A. Knetsch}
\affiliation{Laboratoire d’Optique Appliquée, ENSTA Paris, CNRS, Ecole Polytechnique, Institut Polytechnique de Paris, 91762 Palaiseau, France}%

\author{O. Kononenko}
\affiliation{Laboratoire d’Optique Appliquée, ENSTA Paris, CNRS, Ecole Polytechnique, Institut Polytechnique de Paris, 91762 Palaiseau, France}%

\author{A. Martinez de la Ossa}
\affiliation{Deutsches Elektronen-Synchrotron DESY, Notkestraße 85, 22607 Hamburg, Germany}%

\author{A. Nutter}
\affiliation{Helmholtz-Zentrum Dresden--Rossendorf, Bautzner Landstrasse 400, 01328 Dresden, Germany}%
\affiliation{University of Strathclyde, 107 Rottenrow, Glasgow G4 0NG, United Kingdom}

\author{R. Pausch}
\affiliation{Helmholtz-Zentrum Dresden--Rossendorf, Bautzner Landstrasse 400, 01328 Dresden, Germany}%
\affiliation{Technische Universität Dresden, 01062 Dresden, Germany}

\author{G. Schilling}
\affiliation{Ludwig--Maximilians--Universit{\"a}t M{\"u}nchen, Am Coulombwall 1, 85748 Garching, Germany}%

\author{A. Schletter}
\affiliation{Ludwig--Maximilians--Universit{\"a}t M{\"u}nchen, Am Coulombwall 1, 85748 Garching, Germany}%
\affiliation{Technische Universit{\"a}t M{\"u}nchen, James-Franck-Str. 1, 85748 Garching, Germany}%

\author{S. Sch{\"o}bel}
\affiliation{Helmholtz-Zentrum Dresden--Rossendorf, Bautzner Landstrasse 400, 01328 Dresden, Germany}%
\affiliation{Technische Universität Dresden, 01062 Dresden, Germany}

\author{U. Schramm}
\affiliation{Helmholtz-Zentrum Dresden--Rossendorf, Bautzner Landstrasse 400, 01328 Dresden, Germany}
\affiliation{Technische Universität Dresden, 01062 Dresden, Germany}%

\author{E. Travac}
\affiliation{Ludwig--Maximilians--Universit{\"a}t M{\"u}nchen, Am Coulombwall 1, 85748 Garching, Germany}%

\author{P. Ufer}
\affiliation{Helmholtz-Zentrum Dresden--Rossendorf, Bautzner Landstrasse 400, 01328 Dresden, Germany}%
\affiliation{Technische Universität Dresden, 01062 Dresden, Germany}

\author{S. Karsch}
\email{stefan.karsch@physik.uni-muenchen.de}
\affiliation{Ludwig--Maximilians--Universit{\"a}t M{\"u}nchen, Am Coulombwall 1, 85748 Garching, Germany}%
\affiliation{Max Planck Institut für Quantenoptik, Hans-Kopfermann-Strasse 1, 85748 Garching , Germany}%

\begin{abstract}

We present experimental results on a plasma wakefield accelerator (PWFA) driven by high-current electron beams from a laser wakefield accelerator (LWFA). In this staged setup stable and high quality (low divergence and low energy spread) electron beams are generated at an optically-generated hydrodynamic shock in the PWFA. 
The energy stability of the beams produced by that arrangement in the PWFA stage is comparable to both single-stage laser accelerators and plasma wakefield accelerators driven by conventional accelerators. 
Simulations support that the intrinsic insensitivity of PWFAs to driver energy fluctuations can be exploited to overcome stability limitations of state-of-the-art laser wakefield accelerators when adding a PWFA stage. 
Furthermore, we demonstrate the generation of electron bunches with energy spread and divergence superior to single-stage LWFAs, resulting in bunches with dense phase space and an angular-spectral charge density beyond the initial drive beam parameters. 
These results unambiguously show that staged LWFA-PWFA can help to tailor the electron-beam quality for certain applications and to reduce the influence of fluctuating laser drivers on the electron-beam stability. 
This encourages further development of this new class of staged wakefield acceleration as a viable scheme towards compact, high-quality electron beam sources.
\end{abstract}

\maketitle

\section{Introduction}

Laser wakefield acceleration (LWFA) is a promising high-gradient accelerator technology. It uses intense beams of light to generate strong \textit{wakefields} in a plasma for the acceleration of electrons~\cite{Esarey:2009ks, Malka:2012bi}. 
In LWFA, the ponderomotive force of the laser strongly displaces the plasma electrons from their equilibrium position around the much heavier ions. 
This displacement causes large charge separation fields behind the laser as it traverses the plasma with a velocity close to the speed of light. 
The magnitude of these wakefields is of the order of the cold wavebreaking field $E_0=m_e c \omega_p/e \approx \SI{96}{\giga\volt\per\meter}\times \sqrt{n_e[10^{18}\:\si{\per\cubic\cm}]}$. Here $m_e$ denotes the electron rest mass, $c$ is the speed of light, $\omega_p$ is the plasma frequency, $e$ is the elementary charge and $n_e$ is the plasma electron density. At densities around $10^{18}\:\si{\per\cubic\cm}$ the acceleration gradient in these accelerators is thus several orders of magnitude higher than the breakdown fields in conventional radio-frequency (RF) accelerators ($\sim\SI{100}{\mega\volt\per\meter}$), allowing for a significant downsizing of the accelerator. 
LWFA experiments are routinely performed at numerous high-power laser facilities~\cite{Wenz:2019gc, Oubrerie:2021wi, Kirchen:2021fv, Bloom:2020jb, Schwab:2020fs, Hussein:2019bf, Lemos:2019kd, Gonsalves:2019ht, Constantin:2019jk} and reach high charge ($\sim \si{nC}$)~\cite{Gotzfried:2020da,Couperus:2017bg} combined with an ultra-short bunch duration ($\sim \SI{10}{fs}$)~\cite{Heigoldt:2015cd,Buck:2011dg}, resulting in high peak currents of tens of kA~\cite{Lundh:2011js, Couperus:2017bg}. Furthermore, the bunches typically have a few- micrometer source size at the exit of the accelerator~\cite{Kneip:2010kk, Wenz:2015if, Koehler:2021}, which is paired with few-mrad divergence~\cite{Gotzfried:2020da, Wang:2016gy}. 

One of the most exciting prospective applications of these accelerators is their use for driving a compact free electron laser
(FEL)~\cite{Wang:2021ko}. Similarly, they may form the basis for future compact particle colliders. However,  some limitations of the technology has so far prevented a real breakthrough in these and other applications. Firstly, as a consequence of the very high acceleration gradients~\cite{Dopp:2018kf} and their reliance on non-linear laser propagation~\cite{Streeter:2018eo, Corde:2013gj} small parameter jitters result in large shot-to-shot fluctuations of the electron energy. 
Thus, it is extremely challenging for LWFAs to reach an energy stability and energy spread comparable to conventional RF accelerators. Secondly, the normalized emittance of LWFA electron-beams appears to be limited to around $0.1 - 1$ mm mrad~\cite{Brunetti:2010ue, Weingartner:2012gw, Golovin:2016gl}, likely due to heating of plasma electrons by the intense drive laser, spatio-temporal asymmetries in the driver or the interactions of electrons and the trailing laser fields during acceleration. While significant progress has been made over the past years, including the demonstration of first gain in an LWFA-driven FEL~\cite{Wang:2021ko} and stable long-term operation by actively controlling laser parameters~\cite{Maier:2020kh}, it will remain difficult to solve all of these problems simultaneously. The generation of low-emittance beams will be particularly difficult in proposed multi-stage LWFA concepts for high energy physics~\cite{Lindstrom:2021fh}. In such schemes, the plasma mirrors needed for coupling in multiple laser beams will cause the emittance to increase~\cite{Raj:2020ba}. 

LWFA's particle-driven counterpart, plasma wakefield acceleration (PWFA), relies on the Coulomb field of a relativistic drive-beam and can potentially mitigate some of these problems~\cite{Joshi:2020cu}. In particular, it has been suggested that ultra-low emittance beams can be internally injected into a beam-driven wakefield~\cite{Hidding:2012ep,MartinezdelaOssa:2013hk,MartinezdelaOssa:2015es}. Nonetheless, PWFA remains less common due to its reliance on high-current drive-beams of electrons~\cite{Blumenfeld:2007ja, Litos:2015ke,Lindstrom:2021eo} or protons~\cite{Adli:2018jza}, which were until now only available at a few large-scale accelerator facilities. This situation has changed recently, as we demonstrated that high-current LWFA electron beams are also well-suited to drive strong plasma-waves, even in high-density plasmas~\cite{Gilljohann:2019kc}. These in turn can accelerate witness bunches at gradients of around 100 GeV/m~\cite{Kurz:2021cg}. This possibility opens up a new approach in high-gradient accelerator research, namely using LWFA electron beams to drive a PWFA and to internally inject a high quality beam into the PWFA. While the staging of two plasma-based acceleration methods may sound like an unnecessarily complex approach, we will discuss in this manuscript how the method can efficiently combine the strengths and mitigate the weaknesses of each individual schemes. In particular, our experimental results validate the potential of employing an extra PWFA stage with internal injection to improve the electron quality over the output of a single LWFA. 

The manuscript is structured as follows: 
First we introduce a new all-optical injection scheme for density down-ramp injection in the PWFA stage. 
In a first experiment this new setup is used to investigate the stability of the PWFA stage. We demonstrate stable production of electron-beams from a staged LWFA-PWFA, with the PWFA stage reaching an energy-stability comparable to the drive beam produced in the LWFA stage. Simulations indicate that the intrinsic insensitivity of PWFA to the energy of the drive beam may even allow for using the PWFA stage as a stability transformer, i.e. a system that improves upon the energy stability of a single LWFA stage. In a second experiment, we inject a witness beam in an optically generated density downramp in the PWFA stage to achieve a superior electron beam quality ('quality transformer'). We demonstrate experimentally that its energy spread and divergence beat the respective quantities of its drive bunch, resulting in a net gain in angular-spectral charge density~\footnote{We introduce the angular-spectral charge density as a figure of merit for the beam quality. Its mathematical definition can be found in the supplemental material~\cite{Note3}.}.

\section{Experimental Setup}

\begin{figure*}[t]
  \centering
  \includegraphics*[width=\linewidth]{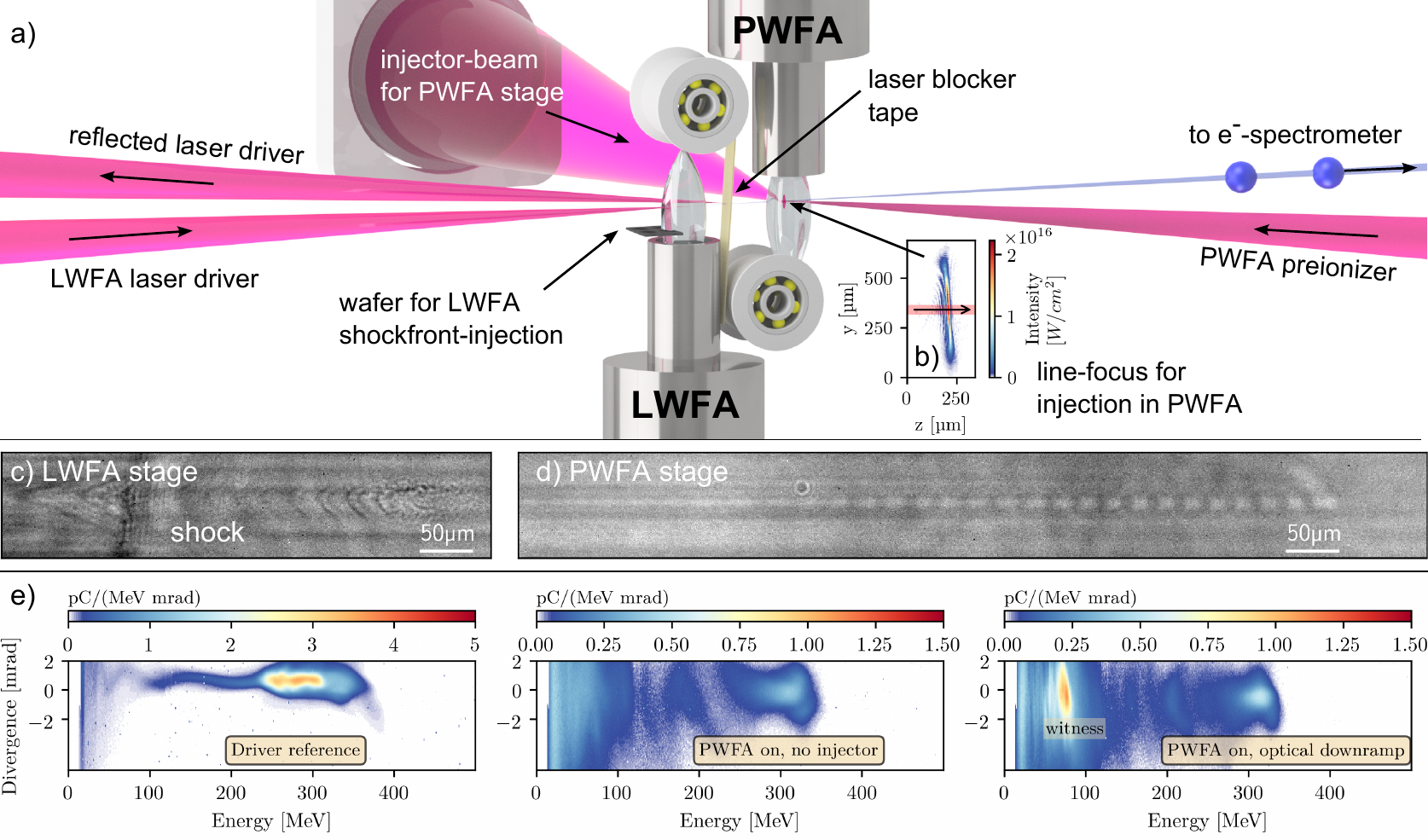}
  \caption{\textbf{Experimental setup for staged LWFA-PWFA with internal injection in the PWFA stage.} Schematic of the double-jet setup for staged LWFA-PWFA experiments. In the first plasma, the drive pulse propagating along the z-axis generates LWFA electrons via shock-front-assisted ionization injection. A $\SI{25}{\micro\meter}$-thick polyimide tape after the first jet prevents the laser from ionizing the second jet for the first set of experiments. An astigmatic laser focus oriented perpendicular to the wakefield axis (b) heats the plasma in the PWFA stage a few ns before the electron driver arrives. A pair of planar plasma density shocks evolves. The second shock provides the downramp for the injection of witness electrons into the PWFA stage. 
  A few-cycle probe was used to image the laser-driven plasma-wave in the LWFA stage (c) and the electron-driven plasma-wave in the PWFA stage (d). Typical spectra (e) of the electron beam from the LWFA stage, the spent drive beam after the PWFA stage without injection and an internally injected witness beam are shown. Without injector laser beam only a broadband background of decelerated LWFA electrons is formed in the PWFA stage. With injector beam the witness is the defined peak on top of this background at around $\SI{70}{MeV}$.}
  \label{experiment1}
\end{figure*}

We have performed a series of closely related experiments on staged LWFA-PWFA. The basic setup is illustrated in Fig.~\ref{experiment1}. To address various aspects of the staging scheme the PWFA parameters are varied. A summary of the PWFA setups presented in this article is given in Table~\ref{PWFA_setup}. 

\textit{Laser.} The laser wakefield accelerator is driven by the ATLAS laser system at the Centre for Advanced Laser Applications in Garching, Germany. During the experiment ATLAS delivered laser pulses with  $ (5\pm 1)\:\si{\joule}$ on target \footnote{For further details on the laser system see supplemental material~\cite{Note3} part \ref{laser}.} with $\SI{30} {fs}$ (FWHM) duration at a central wavelength of $\SI{800} {nm}$. The laser-beam is focused in an f/33 geometry, reaching a peak intensity of $(1.3\pm0.3)\:10^{19}\:\si{\watt\per\cm^2}$, which corresponds to a normalized vector potential of $a_0=2.5$.

\textit{Target.} The laser is focused onto a target consisting of a first and a second gas jets (see. Fig.~\ref{experiment1}a), doubling as the LWFA and PWFA stages, respectively. Both jets are separated by a 10-mm-wide vacuum gap, where diffraction reduces the driver intensity enough to prevent the excitation of any significant wakefield in the PWFA stage. 
Additionally, a tape drive can be inserted between the jets to completely block the laser, which also prevents ionization of the second jet. In this case, an additional low-energy laser beam can be used to preionize the PWFA stage. 

The LWFA stage uses a $\SI{5}{\milli\meter}$ Laval nozzle fed with a 96:4 (molecule-ratio) mixture of hydrogen and nitrogen gas. The 4-mm-long PWFA stage uses either pure hydrogen or mixtures of hydrogen and helium, depending on the specific setup. Both jets'  density profiles are discussed in the supplemental material \footnote{See Supplemental Material at [URL will be inserted by publisher] for additional information on the experimental setup, parameters for simulations and details on data analysis.}. The LWFA is operated at a plateau plasma density of $(1.4\pm 0.1)\: 10^{18}\:\si{\cm^{-3}}$, whereas the PWFA stage is operated at peak densities between $(1-2)\:10^{18}\:\si{\cm^{-3}}$ \footnote{Due to the larger distance of the beam axis from the nozzle, the gas profile has no plateau region}. In the LWFA stage, the hydrodynamic shock-front originating from a silicon wafer edge obstructing the supersonic gas flow triggers shock-front assisted ionization injection~\cite{Thaury:2015dq} (see. Fig.~\ref{experiment1}e) to create the drive bunch for the subsequent PWFA stage.

\textit{Injector Beam.} 
In contrary to previous works using a wire-generated hydrodynamic density down-ramp~\cite{CouperusCabadag:2021gj}, we introduce hydrodynamic-optically field-ionized (HOFI) plasma gradients~\cite{Shalloo:2018fy,Shalloo:2019hv} to facilitate electron injection in the PWFA stage. In this scheme a transversely propagating laser locally ionizes and heats the plasma, forming plasma channel associated with a hydrodynamic shock structure. In contrast other groups' work~\cite{Faure:2010jk,Fourmaux:2012bz,Brijesh:2012cn}, which suffered from alignment sensitivity and pointing jitter, our use of a strongly astigmatic focus (Fig.~\ref{experiment1}b) ensures the formation of two nearly planar shocks perpendicular to the main laser axis. Their large area makes this setup very insensitive to alignment errors and ensures high stability~\footnote{A detailed description of the HOFI shocks can be found in the supplemental material~\cite{Note3} part \ref{supp_injection_methods}.}. 
Our optically induced downramps for injection enable us to tailor the shape, height and gradient of the plasma density down-ramp independently of the gas density or the nozzle geometry.
The relative delay between the injector laser beam and the arrival of the electron beam can be adjusted between $(0-2)\:\si{ns}$. Together with the energy of the injector laser pulse this permits to adjust the parameters of the HOFI-shock~\cite{Shalloo:2018fy} and thus, the plasma-density down-ramp for injection. In our experiment witness bunches are reliably injected at $\SI{1.3}{\nano\second}$ delay and a peak intensity of the injector laser-beam of $\SI{2e16}{\watt\per\cm^2}$ (Fig.~\ref{experiment1}b). 
This added flexibility decouples injection and acceleration in the PWFA stage. In particular, by setting the correct orientation of the astigmatic focus, the shock can be created perpendicular to the beam axis, crucial for the generation of high quality witness beams~\cite{Swanson:2017gn,FanChiang:2020dc} and hard to achieve with supersonic shock formation. In addition, the position jitter of the HOFI injector is only a few $\si{\micro\meter}$ and smaller than typically achieved with wire-generated shocks~\cite{Note3}. 

\textit{Diagnostics.} The main diagnostic in this experiment is a 0.8-m-long dipole spectrometer, placed $\SI{2.9} {m}$ downstream of the target. Electrons are deflected onto a calibrated scintillator screen~\cite{Kurz:2018ji}, whose emission is imaged onto a 12-bit CMOS camera. The spectrometer covers an energy range from $\SI{12} {MeV}$ onward, with transverse angular acceptance range of $\pm\SI{6}{\milli\radian}$.

\begin{figure}[t]
  \centering
  \includegraphics*[width=\linewidth]{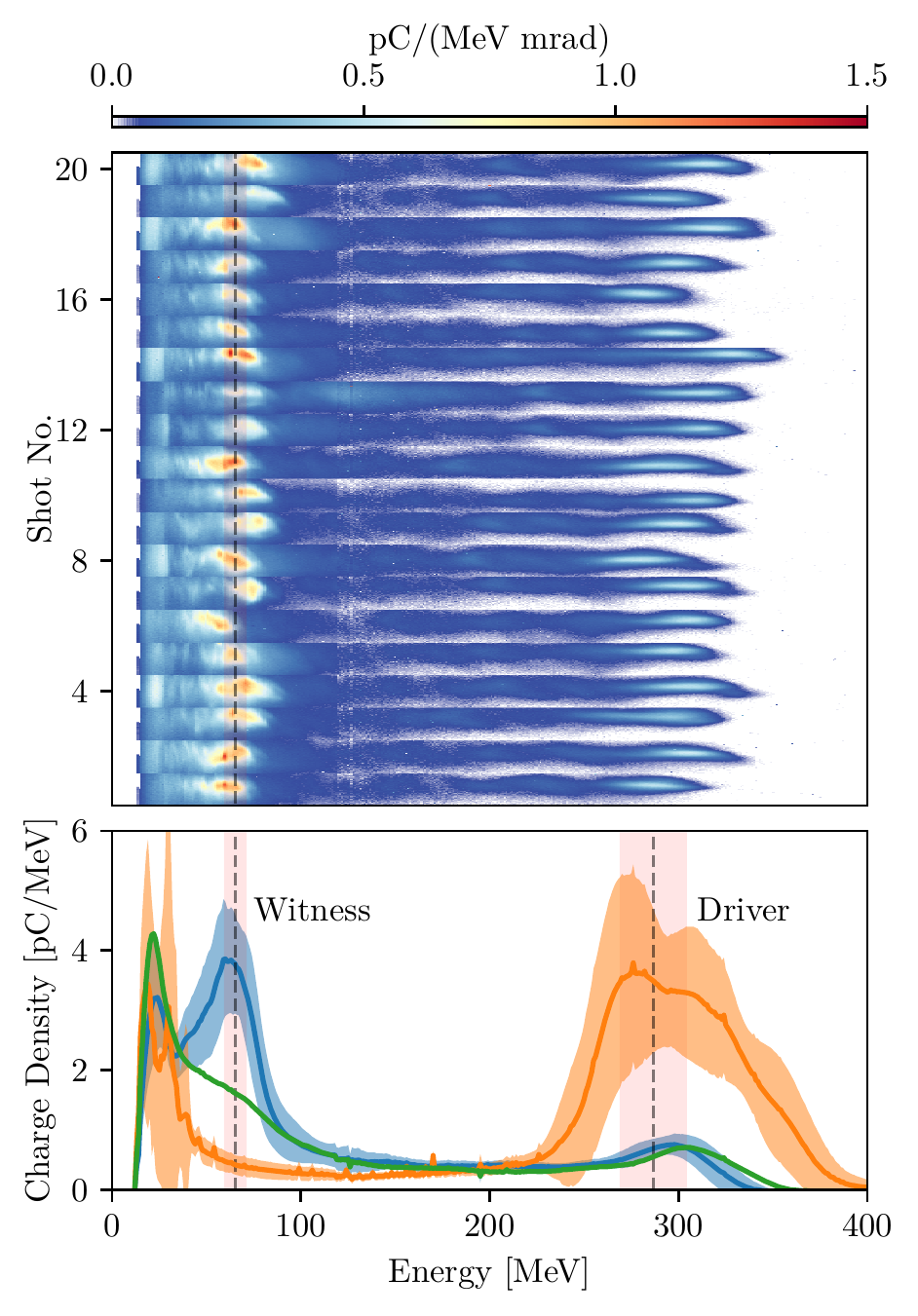}
  \caption{\textbf{Experimental  data  on  stable plasma  wakefield  acceleration}. Top: Output spectra  of 20 consecutive shots with the HOFI-generated shock and pre-ionized plasma in the second jet. Bottom: Electron spectra with injector laser beam (blue). This is compared to spectra without plasma (orange) and with plasma, but no injector (green) in the PWFA stage. The strong charge and energy loss of the driver in the green case and the injection of a high-charge witness in the blue case is evident. Dashed lines and red shaded areas indicate the mean of the energy and its standard deviation of driver and witness bunches respectively. For reference, angle-resolved spectra of the drive beam alone and after the PWFA stage but without injector can be found in Fig.~\ref{experiment1}.} 
  \label{stablerun}
\end{figure}

\section{Stable staged acceleration}

In a first experiment, we investigate the energy stability of the witness injected in the PWFA stage. In this experiment we operate the PWFA stage with a 1:1 mixture of hydrogen and helium (molecule ratio). The laser blocker tape was used to ensure a purely beam-driven wakefield in the second stage.

The LWFA-generated drive bunches in this experiment have a peak energy of $(287\pm18)\:\si{MeV}$ (std). Taking into account all electrons with an energy above $\SI{200}{MeV}$, we measure a charge of $(340\pm46)\:\si{\pico\coulomb}$ (14\%, std), see also Fig.~\ref{experiment1}e (left). Once the beam traverses the second jet (preionizer on, but without injector) the electrons are decelerated and we observe a broadband electron spectrum, see Fig.~\ref{experiment1}e (middle). 

When the injector is activated, a witness bunch is injected at the optically-generated shockfront in the PWFA stage. 
The witness spectra exhibit a distinct energy peak. Furthermore, we observe that the witness injection is very reproducible and, as shown in Fig.~\ref{stablerun}, the energy of the spectral peak fluctuates only within $(65\pm6)\:\si{\MeV}$ ($9\:\%$, std).  
Thus, the absolute fluctuation (red shaded area in Fig.\ref{stablerun}) of the witness peak energy is only one third of the drivers and they are comparable in terms of their relative energy jitter. The charge of the witness beam is $(59\pm 19)\:\si{pC}$ (std).
In terms of energy stability we thus already achieve a performance that is comparable~\cite{Knetsch:2021kj} 
or even superior~\cite{Ullmann:2021ky} to recent experiments on all-optical density downramp (Torch)-injection at large-scale RF-accelerator-driven PWFAs. 

This high stability of the witness energy in our staged LWFA-PWFA appears surprising at first because of the much higher shot-to-shot fluctuations of the LWFA-generated driver as compared to a drive beam from a conventional accelerator. The insensitivity of the witness energy to the driver energy and energy spread can be understood from the expression for the Coulomb field of a highly relativistic ($\gamma \gg 1$), axially symmetric electron beam. Assuming that the beam is contained within a radius $r_0$, the field at a transverse distance $r>r_0$ is given by  
\begin{equation}
\vec{E}_{b}(\zeta, r) = - \frac{I(\zeta)}{2\pi\epsilon_0 c}\frac{\vec{e}_r}{r}  \,,\label{eq:coulomb}
\end{equation}
with $I(\zeta)$ the current profile of the beam in the comoving variable $\zeta = z - ct$, $\epsilon_0$ the vacuum permittivity and $c$ the speed of light. 
We can see from Eq.~(\ref{eq:coulomb}) that the electric field responsible for setting up the plasma-wave is purely oriented in the transverse direction and does not depend on the electron energy.
For sufficiently narrow and high-current beams ($I\gtrsim \SI{1}{kA}$), the Coulomb field strongly expels all plasma electrons from its path leaving behind a homogeneous and symmetric ion column. 
The plasma electrons are attracted back by the space charge field of the ion column, $E_{\rm ion}(r) = -e n_0 r / 2 \epsilon_0$ and start oscillating radially, forming a sheath around the ensuing ion cavities. The maximal radial position of the sheath is usually referred to as the \textit{blowout radius}. 
A useful scaling of the blowout radius $r_{\rm bo}$ can be obtained by calculating the radial distance at which the electrostatic force of the ion background cancels out that of the drive beam. Evaluating Eq.~(\ref{eq:coulomb}) at the point of maximum current $I_0$ we obtain
\begin{equation}
r_{\rm bo} \simeq \sqrt{I_0/\pi e c n_0} \propto \sqrt{I_0/n_0}\,. \label{blowout_radius}
\end{equation}
Note that by balancing electrostatic fields we implicitly restrict the validity of the model to slow plasma electrons in the sheath, for which the Lorentz force is essentially given by the electric field~\footnote{A more careful derivation taking into account the relativistic motion of the plasma electrons and the magnetic field in the blowout~\cite{MartinezdelaOssa:2017} yields $E_{\rm ion}(r) = -e n_0 r / 4 \epsilon_0$ and $r_{\rm bo} = \sqrt{2I_0/\pi e c n_0}$. Thus, the same functional dependence but modified constants.}. 
To estimate the accelerating field we use the notion that, in case of a strong blowout, the plasma sheath approximates a sphere and the longitudinal electric field inside the ion cavity decreases linearly from the cavity center with a slope $\partial_\zeta E_z \simeq e n_0 / 2 \epsilon_0$~\cite{Lotov:2005ed}. 
Thus, evaluating $E_z$ at a distance $r_{\rm bo}$ from the cavity center, we obtain for the maximum accelerating field 
\begin{equation}
E_z^{\rm max} \simeq -e n_0 r_{\rm bo} / 2 \epsilon_0 \propto -\sqrt{I_0 n_0}\,. \label{scaling}
\end{equation}

For small deviations the derived square-root scaling (Eq.~\ref{scaling}) predicts that the relative deviation of the longitudinal electric field is half of the relative deviation of both beam current and plasma density.
To compare the scaling to our experiment it is assumed that the drive bunch length is constant. Thus, the  charge of the drive beam is proportional to its beam current and the 14\% driver charge fluctuation translates into 7\% variation of the witness beam energy. Furthermore, we observe an imperfect regulation of the backing pressure for the PWFA stage leading to a density jitter of $\pm 4  \%$ (std). This translates into 2\% energy jitter of the witness beam energy. Further assuming both error contributions to be independent and normally distributed, we expect an energy jitter of $\SI{5}{MeV}$ for our current setup.
Thus, the prediction of the simplified model on the stability of the staged LWFA-PWFA is consistent with the measured energy of $(65\pm6)\:\si{\MeV}$. 

The ratio of the relative fluctuations of witness energy $\delta E_{\rm witness}$ and driver charge $\delta Q_{\rm driver}$ can be understood as a measure for the resilience of the PWFA stage to variations of the driver. The measured value of $\left|\delta E_{\rm witness}[\%]\right|= \leq 0.68 \left|\delta Q_{\rm driver}[\%]\right|$~\footnote{This number includes all contributions to the variation of witness energy. The '=' sign would apply if the variation in witness energy is only due to charge variations of the driver.} which is smaller than 1 indicates a damping behaviour.

\begin{figure*}[t]
  \centering
  \includegraphics*[width=\linewidth]{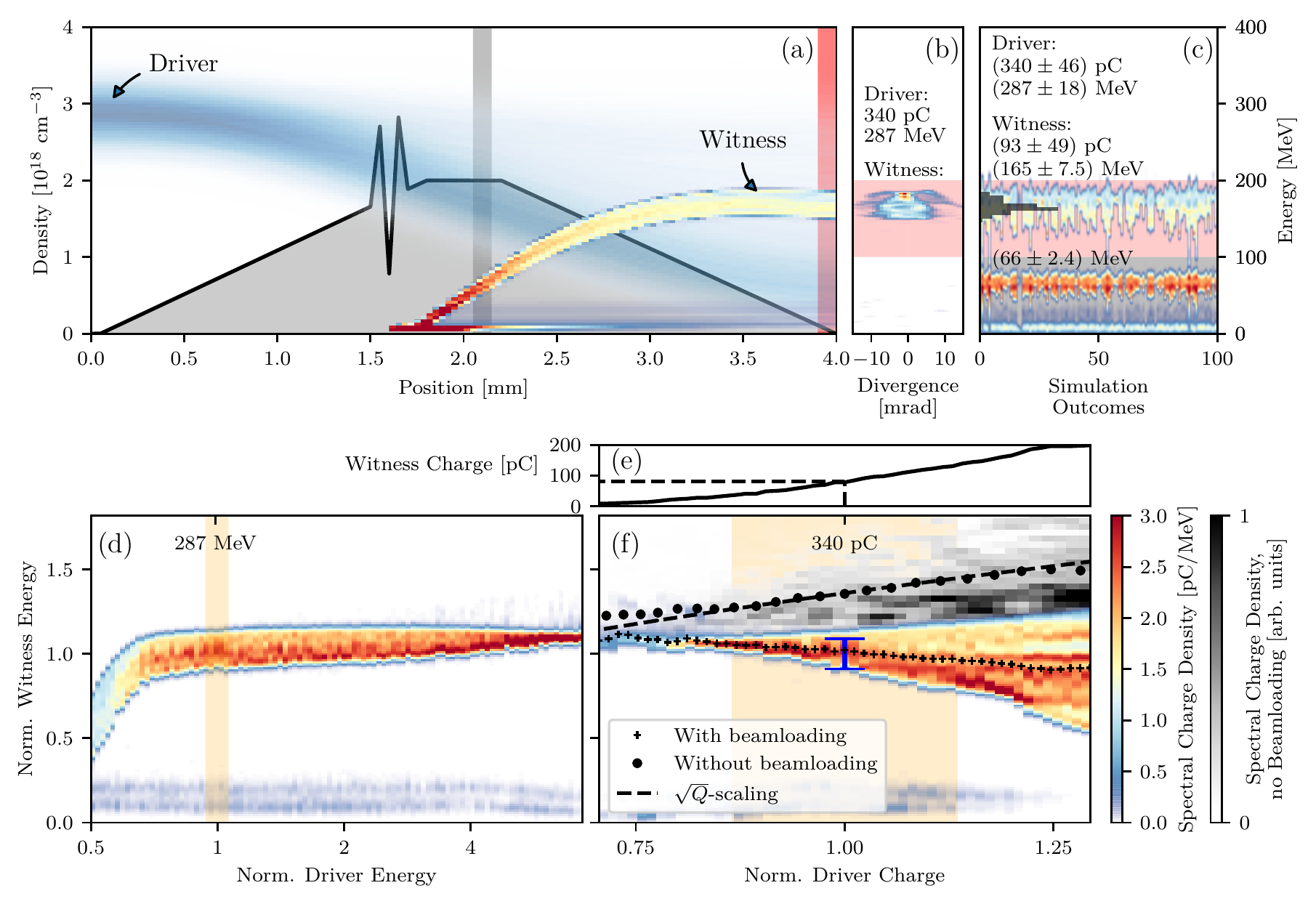}
  \caption{\textbf{PIC simulations on stable staged acceleration.} (a) Modelled density distribution and evolution of driver and witness energy in the PWFA stage. (b) Angular resolved witness spectrum for the experimental driver parameters. (c) Set of 100 simulations with randomized variation of driver charge and energy with standard deviation as in the experiment. For comparison the spectra of an earlier simulation step (gray shaded in (a)) are shown where the energy gain resembles the experimental outcome. (d) and (f) show the spectral charge density of the witness for parameter scans of the driver energy and charge. Given that driver depletion can be neglected (here driver energy $>\SI{200}{MeV}$) the witness energy does barely depend on the driver energy. (f) shows the spectral charge density of the witness bunch (color scale and black crosses) as a function of the driver charge. With increasing driver charge the witness charge increases (e), the witness spectrum broadens and its mean energy slightly decreases. For comparison the energy gain of a test bunch in an unloaded wakefield with equal driver is plotted (gray scale and black dots). In both scans the experimental working point is indicated by the orange shaded area. The simulation outcome in plots (d) and (f) is normalized to the experimental working point in (b). For comparison the experimental standard deviation of the witness energy is plotted in blue. Note that the experimental result is normalized independently. }
  \label{simulations}
\end{figure*}

In future experiments, the intrinsic insensitivity of the PWFA stage to the energy of the drive beam may even permit to increase the energy stability of a staged LWFA-PWFA beyond that of the LWFA alone. We have modeled the scenario of density downramp injection in a plasma wakefield accelerator stage using the quasi-3D particle-in-cell code FBPIC~\cite{Lehe:2016dn}. As shown in Fig.~\ref{simulations}a, the input parameters are similar to our experiment (for details see supplemental material~\cite{Note3}). 

In a first set of 100 simulations we model the experimentally observed drive bunch in terms of average charge and energy and their random variations (mean and standard deviation as in the experiment). As shown in Fig.~\ref{simulations}c the mean energy of the witness bunch is $165\pm7\si{MeV}$~(std), and $\left|\delta E_{\rm witness}[\%]\right| \leq 0.34 \left|\delta Q_{\rm driver}[\%]\right| $. Thus, the relative energy variation of the simulated witness is
smaller than in the experiment. 

We study the reason for this finding in independent scans of driver energy and charge. First, we vary the driver energy between half and eight times the value of the experiment, while keeping the driver charge constant at $\SI{340}{pC}$. Beyond a certain energy level, which is on the order of $\SI{200}{MeV}$ for our simulation parameters, depletion of the driver can be neglected (see Fig.~\ref{simulations}d). As follows from \ref{eq:coulomb}, the wakefield strength does not depend on the driver's electron energy and thus the witness energy stays constant. This finding holds for a broad driver energy range that far exceeds the measured energy fluctuations in our experiment (the latter are highlighted in orange).

In a second set of simulations we scan the driver charge between $240-440~\si{pC}$ while keeping the energy constant at $\SI{287}{MeV}$. In these simulations the mean witness energy slightly decreases with increasing driver charge around the experimental working point ($\delta E_{\rm witness}[\%] =-0.35\cdot \delta Q_{\rm driver}[\%]$). Furthermore, the spectrum broadens towards stronger drivers. This downshift and broadening of the spectra can be explained with beam loading of the wakefield. In the simulations the injected charge is positively correlated to the driver charge (Fig.~\ref{simulations}e). Thus, the amount of injected charge can attenuate, or even overcompensate the higher energy gain expected for stronger drive beams in the PWFA via beam loading. 

To quantify this effect we include a species of test-particles in the PIC simulations to sample the longitudinal phase space. We then compare the energy gain of the witness in the case of a beam-loaded and an unloaded wakefield (method described in the supplemental material~\cite{Note3}). In Fig.~\ref{simulations}f both cases are compared. The mean witness energy in the hypothetical unloaded case roughly follows the $\sqrt{Q}$-scaling for the longitudinal wakefield strength as a function of the driver charge ($\delta E_{\rm witness}[\%] =+0.48\cdot\delta Q_{\rm driver}[\%]$). 
In the experiment the injected witness charge is expected to fluctuate not only as a function of the driver charge, but also due to other parameters that were kept constant in the simulations (e.g. gas density distribution, down-ramp gradient and height). Thus, a random contribution to the witness energy, depending on the magnitude of the additional witness charge variation, is added and $\left|\delta E_{\rm witness}/\delta Q_{\rm driver}\right|$ must be expected to be higher than in the simulations. 
Note that the sign of $\delta E_{\rm witness}/\delta Q_{\rm driver}$ is not experimentally observable because the PWFA stage deletes the information about the initial driver charge. Thus, only mean and standard deviation of driver and witness charge (orange shade area and blue bar in Fig.~\ref{simulations}f) of similar runs can be compared and correlations as in Fig.~\ref{simulations}f can not be revealed in the experiment.

The mean energy of the witness beam in the simulations is 2.5 times higher as compared to the experiment. While this experiment was not optimized for highest energy, but highest stability, the main reason for this discrepancy might be that the wakefield strength is overestimated in our simulations. The drive bunch is initiated without taking the interaction at the laser blocker tape into account. Simulations showing the influence of the driver emittance on the witness beam and a discussion of the angular distribution of the witness beam can be found in the supplemental material~\cite{Note3}.  

The beams presented in Fig. \ref{stablerun} do not only exhibit high energy stability, but they also carry a significant fraction of the energy of the LWFA-generated drive bunch. We calculate the overall energy transfer efficiency as the ratio of the integrated energy of the incident LWFA-generated driver bunch ($E_{\text{driver}}$) and energy gain of the PWFA witness bunch ($\Delta E_{\text{witness}}$) according to following definition~\footnote{Note that e.g.~\cite{Litos:2015ke,Lindstrom:2021eo} use a different definition for the energy transfer efficiency, namely the driver's energy loss $\Delta E_{\text{driver}}$ instead of the initial driver energy in the denominator ($\tilde \eta = {\Delta E_{\text{witness}}}/{\Delta E_{\text{driver}}}$). Their definition thus yields high efficiency figures even far from driver depletion.} 
\begin{equation}
    \eta = \frac{\Delta E_{\text{witness}}}{E_{\text{driver}}}, 
\end{equation}
 For data from Fig.~\ref{stablerun}, we find $E_{\text{driver}}=(102\pm14)\si{mJ}$ and $\Delta E_{\text{witness}}=(4\pm1)\si{mJ}$, which thus yields an overall efficiency from incident driver to witness of $\eta=4\%$ and, for some experimental conditions \footnote{This efficiency was measured in sets with less stable witness formation. A representative shot is presented in the supplemental material~\cite{Note3}.}, this efficiency reaches up to 10 percent. This is at least a factor of 2 more than shown in either experiments with external~\cite{Litos:2016, Litos:2015ke, Lindstrom:2021eo} or internal~\cite{Knetsch:2021kj,Ullmann:2021ky,Kurz:2021cg} injection and, to our knowledge, the highest total driver to witness energy transfer efficiency observed for a PWFA to date. 
 
 As we will discuss in the following section, the hybrid approach not only reaches a substantial energy transfer efficiency, but can also lead to a net improvement of selected beam-parameters.

\section{High quality electron beams from staged acceleration}
As discussed, a staged LWFA-PWFA helps to decouple the electron energy from 
shot-to-shot fluctuations of the driver bunch. 
The established stability of the wakefield in combination with a controlled injection enables the pursuit of a beam-quality transformer~\cite{MartinezdelaOssa:2019bm}.
As a figure of merit we use the angular-spectral charge density, i.e. the charge per solid angle ("angular") and energy interval ("spectral"). We define it as the spectrally resolved charge within the RMS-divergence divided by the solid angle corresponding to this divergence \footnote{A more detailed definition of the angular-spectral charge density and a description of the analysis can be found in the supplemental material~\cite{Note3}.}. To deduce the solid opening angle rotational symmetry of the witness bunches is assumed. Experimental evidence for an overall improvement regarding the electron-beam density is presented in the subsequent section. 

\begin{figure*}[t]
  \centering
  \includegraphics*[width=\linewidth]{ 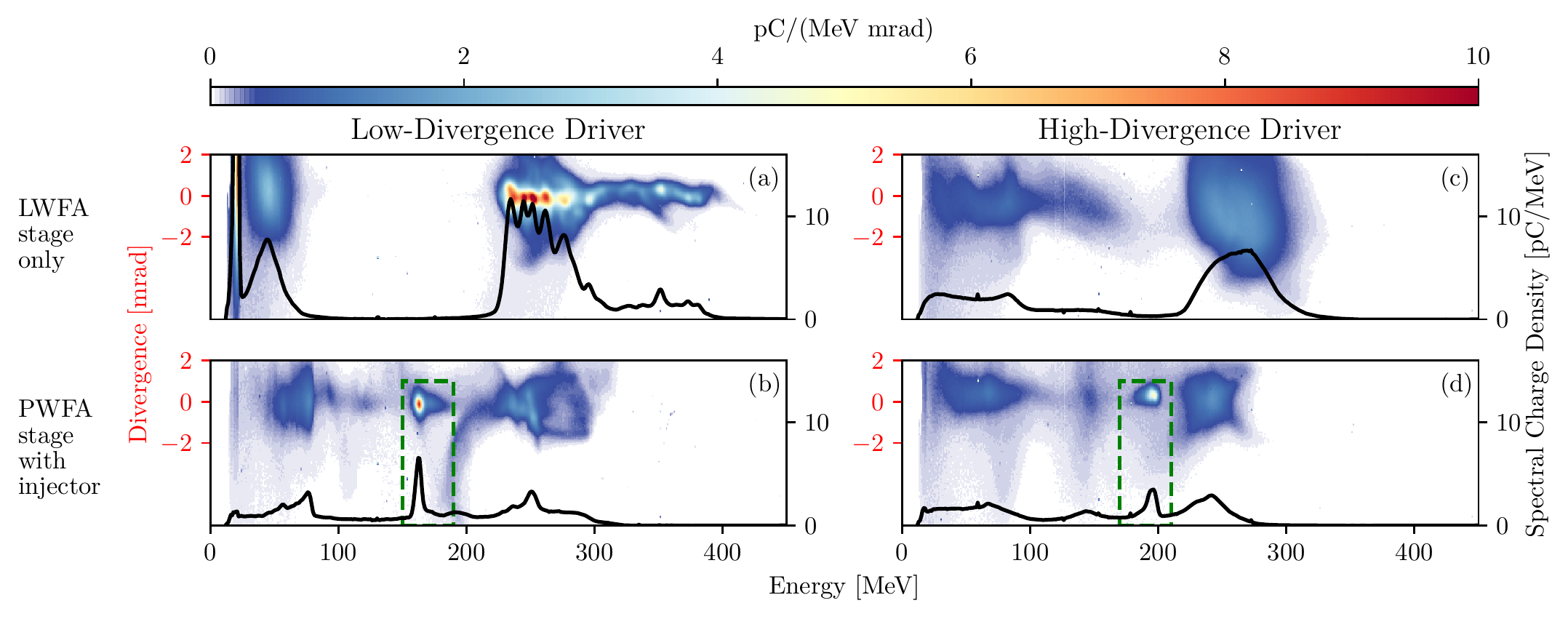}
\caption{\textbf{Increase of angular-spectral charge density in the PWFA stage}. (a) Typical spectrum of a low divergence LWFA-generated drive beam with $\SI{640}{\pico\coulomb}$ bunch charge in the high energy feature and an average divergence of $\SI{0.41}{\milli\radian}$ (RMS of super-Gaussian fit), leading to an angular-spectral charge density of $\SI{5}{\pico\coulomb\per(\mega e\volt\:\micro\steradian)}$. (b) Spectrum after the PWFA stage with optimized beamloading for high charge density of the witness beam. Due to the lower divergence of $\SI{0.28}{\milli\radian}$ (RMS of super-Gaussian fit) of the $\SI{30} {pC}$ witness beam its angular-spectral charge density is 40\% higher than the driver ($\SI{7}{\pico\coulomb\per(\mega e\volt\:\micro\steradian)}$). (c) Typical LWFA driver spectrum for the high divergence case ($\SI{1.2}{\milli\radian}$, RMS of super-Gaussian fit). Using this beam with a charge of $\SI{400}{pC}$ and an angular-spectral charge density of only $\SI{0.4}{\pico\coulomb\per(\mega e\volt\:\micro\steradian)}$, a witness beam (d) with  $\SI{0.22}{\milli\radian}$ RMS divergence, $\SI{20}{\pico\coulomb}$ charge and an angular-spectral charge density of $\SI{6}{\pico\coulomb\per(\mega e\volt\:\micro\steradian)}$ is generated.} 
  \label{angular_spectral_density}
\end{figure*}

For this demonstration, the laser blocker tape between both acceleration stages was removed. By that we avoid additional emittance growth of the drive beam due to the Weibel instability~\cite{Raj:2020ba}. The resulting lower emittance of the driver results in a denser drive bunch in the PWFA stage. These unperturbed drive beams are expected to generate a stronger and more symmetric blowout, which is crucial for the generation of high quality witness beams~\cite{Zhang:2019}. Similar to previous experiments~\cite{Gotzfried:2020da, Kurz:2021cg},
in order to exclude the laser beam as the dominating driver of the second stage a distance $\geq 1 \si{\centi\meter}$ between both jets is chosen. Figure \ref{angular_spectral_density}a shows a typical drive bunch generated by our 150-TW-LWFA for this set of experiments. The average charge in a set of 30 shots was $(657\pm61)\:\si{\pico\coulomb}$ (std) in the high energy feature at $\SI{250}{MeV}$. The shot shown in Figure \ref{angular_spectral_density} has a charge of $\SI{640}{\pico\coulomb}$ in the high energy feature and its average divergence is $0.41 \:\si{\milli\radian}$ (RMS of super-Gaussian fit, for details see supplemental material~\cite{Note3}). 
The angular-spectral charge density is $5\: \si{\pico\coulomb\per(\mega e\volt\:\micro\steradian)}$~\footnote{Please consult the supplemental material~\cite{Note3} for a detailed discussion of the influence of the jet separation, different injection conditions and low-energy features in the spectra.}.

In our experiment, we are able to modify the density down-ramp for injection in the PWFA stage by tuning the delay, intensity and position of the injector laser pulse and thus independently of the gas density, gas species and nozzle profile. 
With the mrad-level divergence of the LWFA-generated bunch, the distance between LWFA and PWFA stage serves as a parameter to adjust the density of the drive bunch when entering the second stage. Thus, the drive beam evolution and consequently the strength of the wakefield at the time of injection can be controlled.  
This set of free parameters is used to optimize the injected charge and in particular the angular-spectral charge density of the witness bunches from the PWFA stage. 

For an injected witness charge of about $\SI{30}{\pico\coulomb}$ there is a regime where a flattening of the longitudinal phase-space is observed. This manifests itself in a reduced energy spread and an increased spectral charge density of the witness. The charge-dependent behaviour of the witness' spectral charge density hints at beamloading as an explanation for this observation~\footnote{A comparison of the spectral charge density for different amount of injected charge can be found in the supplemental material~\cite{Note3}}. Witness electron bunches with low divergence, low energy spread and high spectral charge density are produced in a fraction of the shots. The reproducibility of such beams is limited by the shot-to-shot fluctuations of the injected witness charge. 
 Fig.~\ref{angular_spectral_density}b depicts an example for an separation of LWFA and PWFA of $\SI{19}{\milli\meter}$. The peak energy of this witness bunch is $\SI{162}{MeV}$. The bunch charge of the narrow-band bunch is $(31\pm5)\:\si{\pico\coulomb}$. Its FWHM and RMS (from Gaussian fit to spectrum) energy spread is $\SI{5.6}{MeV}$ and $\SI{2.4}{MeV}$ respectively, approaching the energy resolution of the non-imaging dipole spectrometer \footnote{In our specific dipole spectrometer setup a mono-energetic electron bunch at 162 MeV with a FWHM divergence of 0.6 mrad appears to have an energy spread of $\SI{3.5}{MeV}$ (FWHM).}. The beams thus exhibit a very good, low energy-spread-to-gain ratio of $3.5\%$, commonly defined as the FWHM-energy-spread of the witness divided by its energy gain. This is 5 times less than the relative energy spread of the driver in this experiment (18\%).
 
 Furthermore, we observe that these witness bunches have an average divergence of $\SI{0.28}{\milli\radian}$ (RMS of super-Gaussian fit) ($\SI{0.6}{\milli\radian}$ FWHM), only. Using pure LWFA, similarly small divergences were only observed for near-GeV beams~\cite{Ke:2021hh}. Since the beam divergence is given by the ratio of transverse to longitudinal momentum, this hints at a competitively small beam emittance as we will elaborate on later. Combined with its charge of $\SI{30}{\pico\coulomb}$, this yields an angular-spectral charge density of $\SI{7}{\pico\coulomb\per(\mega e\volt\:\micro\steradian)}$, which is approximately 40\% denser than the drive beam. As seen both in simulations and in first experimental results already a slight improvement in terms of electron beam quality can enable further progress in realizing free electron lasers~\cite{Lehe:2014vu,Couprie:2015te,Andre:2018kr,Maier:2012eq,Wang:2021ko}. Thus, the beams generated in our PWFA are very promising for various such applications. 
 In particular, with a divergence after extraction of well below $\SI{1}{mrad}$ in combination with \%-level energy spread, the beams can be coupled into a beamline and transported without significant degradation \footnote{Beam degradation due to chromatic beam transport can alternatively be described as a growth of normalized emittance during free-space propagation. Following the description by~\citet{Migliorati:2013jc}, the normalized emittance evolution of a beam with energy spread and divergence is given by 
$\epsilon_{\text{n,rms}}=\gamma\sqrt{\sigma_\gamma^2\sigma_{x'}^4(z-z_0)^2+\epsilon_{\text{rms}}}$. The emittance growth for our beam parameters evaluates to $\SI{0.30}{\milli\meter\:\milli\radian}$ per $\SI{1}{m}$ of free-space propagation.}.

Remarkably, the production of these dense, low-divergence witness beams is not limited to highly optimized, sub-mrad drive beams such as the one shown in Fig.~\ref{angular_spectral_density}a, but is also seen in experiments with significantly more divergent LWFA-beams as driver. Figure \ref{angular_spectral_density}c shows a representative shot from a different data-set with the drive bunch containing $\SI{400} {\pico\coulomb}$, a divergence of $\SI{1.2}{\milli\radian}$ (RMS of super-Gaussian fit) and thus with a much lower angular-spectral charge density of $\SI{0.4}{\pico\coulomb\per(\mega e\volt\:\micro\steradian)}$ at $\SI{270}{\mega e\volt}$. These beams can still drive a plasma wakefield and, as shown in Fig.~\ref{angular_spectral_density}d, yield high-quality witness bunches with a similarly small divergence of $\SI{0.22}{\milli\radian}$ (RMS of super-Gaussian fit) and $2.3\%$ (RMS of Gaussian fit) energy spread. At a charge of $\SI{20}{\pico\coulomb}$ the angular-spectral charge density of these witness bunches evaluates to $\SI{6}{\pico\coulomb\per(\mega e\volt\:\micro\steradian)}$ at $\SI{195}{\mega e\volt}$. This is an order of magnitude denser than the driver. The witness properties, in particular its angular-spectral charge density, thus appear to be largely insensitive to the driver divergence in an interval spanning more than one order of magnitude~\footnote{Dedicated simulations to investigate the influence of the driver divergence and emittance can be found in the supplemental material~\cite{Note3}.}. 
The similar, small divergence of the witness beams in both scenarios indicates that the injected electrons mainly carry the intrinsic transverse momentum spread of our injection method and are little affected by either the electron driver or the remainder of the laser pulse from the LWFA stage. In the following we will establish reasonable upper and lower limits on the emittance of the witness beam. 

Shock-injected electrons originate from the bubble sheath, and therefore have previously been transversely displaced by the driver. An upper limit for the divergence and emittance of the witness beam in this scheme can thus be calculated by the transverse momentum of the sheath electrons falling back onto axis at the rear of the bubble.

\begin{table*}
\caption{\textbf{Summary of experimental conditions.}} 
\begin{tabular}{ |l|l|l|l|l| } 
 \hline
Figure& \ref{experiment1} and \ref{stablerun} & \ref{angular_spectral_density}a-b and \ref{angular_spectral_density_supp} & \ref{angular_spectral_density}c-d \\
\hline
Driver charge & $(340 \pm 46)\:\si{\pico\coulomb}$ & $(657\pm61)\:\si{\pico\coulomb}$ & $(461\pm 99)\:\si{\pico\coulomb}$\\
Driver energy& $(287\pm18)\:\si{MeV}$ & $(235\pm14)\:\si{MeV}$ & $(284\pm 30)\:\si{MeV}$\\
Driver divergence (FWHM,  & $(1.1\pm 0.2)\:\si{\milli\radian}$&  $(0.9\pm0.1)\:\si{\milli\radian}$ & $(4.4\pm 0.5)\:\si{\milli\radian}$ \\

\hspace{.3cm}no laser blocker, no 2nd jet) & & & \\
Gap between stages& $1\:\si{\centi\meter}$ & $(1-1.9)\:\si{\centi\meter}$ & $1.25\:\si{\centi\meter}$\\
Nozzle PWFA& $4\:\si{\milli\meter}$ & $7\:\si{\milli\meter}$ & $4\:\si{\milli\meter}$\\
Gas PWFA& 50\% $\text{H}_2$ + 50\% He & $\text{H}_2$ & $\text{H}_2$\\
Density PWFA & $(2.0\pm0.2) 10^{18}\:\si{\cm^{-3}}$ & $(1.1\pm0.2)) 10^{18}\:\si{\cm^{-3}}$ & $(2.0\pm0.2) 10^{18}\:\si{\cm^{-3}}$ \\
Down-ramp generation & optically & optically & wire\\
Laser blocker tape&  $25\:\si{\micro\meter}$ Kapton & no & no\\
Preionizer& on & off & off\\
\hline
\end{tabular}
\label{PWFA_setup}
\end{table*}

We estimate the order of magnitude of the intrinsic transverse momentum in our implementation of density down-ramp injection based on the simplified model derived above. 
From the transverse momentum betatron trajectories and the normalized emittance of the electron bunch are calculated~\cite{Note3}.  
For a driver current of $\SI{20}{\kilo\ampere}$, a plasma density of $n_0=1\times10^{18}\:\si{\centi\meter^{-3}}$ and a Lorentz factor of $\gamma=300$ the upper limit for the divergence angle at the end of the longitudinal acceleration is $\sigma_\theta=\SI{4}{\milli\radian}$. At this point the betatron amplitude of the electrons defining the contour of the phase-space ellipse is $\sigma_x=\sigma_\theta c /\omega_\beta=
\SI{0.5}{\micro\meter}$. Here $\omega_\beta=\omega_p/\sqrt{2\gamma}$ is the local betatron frequency and $\omega_p=\sqrt{e^2 n_0/\epsilon_0 m_e}$ the plasma frequency. 
This numbers yield an upper limit for the normalized emittance of 
\begin{equation}
\epsilon_{n}< \gamma\sigma_\theta^{\text{max}}\sigma_x^{\text{max}} =0.6\:\si{\milli\meter\:\milli\radian}.
\end{equation} 
The estimated divergence value of $\SI{4}{\milli\radian}$ is more than 10 times larger than the measured divergence of the witness beam. This observation indicates a considerably smaller emittance due to less transverse momentum of the electrons at the position and time of injection. A damping of the transverse momentum can happen because of a transversely defocusing field of the on-axis density spike at the rear of the bubble~\cite{Xu:2017gt} or because of the space-charge field of the injected electrons themselves effectively lowering the focusing fields inside the bubble while being injected~\cite{MartinezdelaOssa:2017}. 

A lower limit for the normalized emittance can be calculated from the measured free-space divergence, assuming that the transverse momentum is identical inside the wakefield and after extraction to free-space. However, adiabatic matching of the witness beam divergence may occur in the density down-ramp of the jet or in a possible transition from a blowout to a linear wakefield at the end of the acceleration process~\cite{Mehrling:2012ky, Floettmann:2014dl,Ariniello:2019cg}. Thus, just assuming the measured free-space divergence to be indicative for the transverse momentum during the acceleration will likely underestimate the emittance~\footnote{Such a calculation yields a rms radius of the betatron motion of $\sigma_x^{\text{min}} = \SI{0.04}{\micro \meter}$ and a normalized emittance of $\epsilon_{n}>\gamma\sigma_\theta^{\text{min}}\sigma_x^{\text{min}} = \SI{0.004}{\milli\meter\:\milli\radian}$}. We can also compare the theoretical estimates to high-resolution PIC simulations of downramp injection from the previous section, cf. Fig.~\ref{simulations}b. While these simulations only roughly approximate our experimental conditions, the emittance of the high density part of the witness~\cite{Note3} fits well into our estimates with $\epsilon_{n}^{sim}\approx \SI{0.25}{\milli\meter\:\milli\radian}$. 
While these estimates hint at a small witness emittance that is independent from and lower than the driver's, additional diagnostics and measurements will be required to determine its actual value.

\section{Conclusion \& Outlook}

In the present paper we provide first evidence that combined LWFA-PWFA offers a path to generate witness beams with improved quality parameters as compared to a single-stage LWFA. 
We have investigated the energy stability of electron acceleration in a staged LWFA-PWFA with an optically-induced density down-ramp in the PWFA. The energy of the witness beam is largely insensitive to the energy, energy spread, and emittance of the drive bunch produced in the LWFA stage. 
Furthermore, the addition of a subsequent PWFA stage for injection and acceleration makes use of the intrinsic resilience of beam-driven wakefields to shot-to-shot variations of the drive beam charge. 
This behaviour is contrary to LWFA, where the electron properties are strongly correlated to variations of the driving laser pulse energy and focus position~\cite{Maier:2020kh}. In our staged scheme we observe similar shot-to-shot stability as in PWFA experiments driven by conventional RF accelerators, despite substantially stronger fluctuations of the driver in terms of charge and energy. Simulations show that in our hybrid scheme, the PWFA stage can generate electron-beams with higher stability than the driving LWFA itself. 
Our simulations suggest that the stability of the witness energy can be increased even beyond the stability of an unloaded wakefield when controlling the amount of injected charge. For this more stable targets (e.g. gas cells) should be employed. Furthermore, the injection and acceleration need to be further decoupled to better control the amount of injected charge (e.g in a plasma photo-cathode scheme).

The position of the optically-generated density down-ramp in the PWFA stage is very stable and thus one major source of witness energy fluctuations is eliminated. Furthermore, injection at such planar optically-generated shocks and acceleration in the PWFA stage is very robust against pointing fluctuations of the laser driver and as a result the LWFA electron-beam. This is because both the jet dimension and the transverse extent of the astigmatic focus of the injector laser beam in the PWFA stage are much larger than the typical transverse jitter of the drive laser. 
The presented laser blocker-free setup is essentially self-aligning because the spent laser driver acting as an ionizer for the PWFA stage always propagates sufficiently collinearly with the electron-beam. 

Without the laser blocker and by controlling the amount of injected witness charge we achieve narrow-band witness spectra via beam-loading. 
The angular-spectral charge density of PWFA-beams injected at an optically induced density down ramp exceeds the one of the drive beam being used. 
We thus demonstrate that an additional PWFA stage with internal injection driven by a LWFA acts as a beam-quality transformer. 

The energy transfer efficiency is higher than in previous PWFA-experiments. The ratio of the integrated energy of the witness to the integrated energy of the incident drive bunch is up to 10\%. Furthermore, a high beam quality of the witness beam and simultaneously a high energy gain of 65 \% of the initial electron-energy of the driver has been shown. 
However, there are different limiting factors for achieving a witness energy that exceeds the driver energy in our specific experimental implementation of the PWFA stage. As seen in Fig.~\ref{simulations}a, the effective acceleration length is limited to $\sim\SI{1.5}{mm}$ in our first set of experiments, because of the long density downramp and the associated dephasing of the witness bunch in the PWFA target. Also, under our experimental conditions the injected witness charge should be limited to a few $\SI{10}{pC}$ to avoid strong beam loading, which limits the energy gain in the PWFA. Furthermore, due to the free space drift between LWFA and PWFA the drive bunch is not matched into the PWFA plasma. Thus, the emittance of the drive beam degraded when entering the PWFA~\cite{Note3} and its full ability to drive strong wakefields is not exploited. This can be addressed by implementing a low density passive plasma lens between both stages. Or at least mitigated by a reduction in stage separation, which in the case without blocker tape comes at the expense of a stronger remaining laser driver from the LWFA. 

Due to the low free-space divergence and low energy-spread of our witness beams their emittance growth during propagation in free-space is smaller than for most beams reported in previous wakefield accelerator experiments. Thus, they are suited for applications involving an electron beam transport line. In follow-up experiments a careful and full assessment of the overall and slice emittance has to be done to asses the suitability of these beams for a free-electron laser. 
Furthermore, we plan to implement advanced injection schemes such as the plasma photocathode~\cite{Hidding:2012ep} or wake-induced ionization injection~\cite{MartinezdelaOssa:2013hk,MartinezdelaOssa:2015es} that promise a further reduction of the witness emittance. 

The two-stage LWFA-PWFA scheme is particularly interesting for facilities offering 100-TW to PW-scale laser power that can generate electron beams with nC-class beam charge and tens of kA peak current. Energy transfer efficiency and electron-energy gain shown in this paper encourage to consider a final PWFA stage with internal witness injection as a beam quality and stability booster after a single or multiple LWFA stages. In the latter case this would relax the demands on emittance preservation in the LWFAs by far. This scheme may be a promising future research direction for high energy physics applications of wakefield accelerators. 

\vspace{1cm}
\emph{Acknowledgements.} This work was supported by the DFG through the Cluster of Excellence Munich-Centre for Advanced Photonics (MAP EXC 158), TR-18 funding schemes and the Max Planck Society. Furthermore, this work has been carried out within the framework of the EUROfusion Consortium and has received funding from the Euratom research and training programme 2014-2018 and 2019-2020 under Grant agreement No. 633053. F.M.F.\ is part of the Max Planck School of Photonics supported by BMBF, Max Planck Society, and Fraunhofer Society. S.C., M.G., A.K., and O.K. were supported by the European Research Council (ERC) under the European Union’s Horizon 2020 research and innovation programme (Miniature beam-driven Plasma Accelerators project, ERC Grant Agreement No. 715807).

\emph{Author contributions.} F.M.F., A.D., F.H., K.v.G., F.I., G.S., A.S., E.T. and S.K. set up and performed the experiment. F.M.F. analyzed the experimental data. A.D. performed simulations. F.M.F. and A.D. wrote the manuscript with input from all co-authors. S.K. supervised the project.

\bibliography{references.bib}

\end{document}


\title{Supplemental Material: Stable and high quality electron beams from staged laser and plasma wakefield accelerators}

\author{F. M. Foerster}
\email{moritz.foerster@physik.uni-muenchen.de}
\affiliation{Ludwig--Maximilians--Universit{\"a}t M{\"u}nchen, Am Coulombwall 1, 85748 Garching, Germany}%

\author{A. D{\"o}pp}
\affiliation{Ludwig--Maximilians--Universit{\"a}t M{\"u}nchen, Am Coulombwall 1, 85748 Garching, Germany}%
\affiliation{Max Planck Institut für Quantenoptik, Hans-Kopfermann-Strasse 1, 85748 Garching , Germany}%

\author{F. Haberstroh}
\affiliation{Ludwig--Maximilians--Universit{\"a}t M{\"u}nchen, Am Coulombwall 1, 85748 Garching, Germany}%

\author{K. v. Grafenstein}
\affiliation{Ludwig--Maximilians--Universit{\"a}t M{\"u}nchen, Am Coulombwall 1, 85748 Garching, Germany}%

\author{D. Campbell}
\affiliation{Ludwig--Maximilians--Universit{\"a}t M{\"u}nchen, Am Coulombwall 1, 85748 Garching, Germany}%
\affiliation{University of Strathclyde, 107 Rottenrow, Glasgow G4 0NG, United Kingdom}

\author{Y.-Y. Chang}
\affiliation{Helmholtz-Zentrum Dresden--Rossendorf, Bautzner Landstrasse 400, 01328 Dresden, Germany}%

\author{S. Corde}
\affiliation{Laboratoire d’Optique Appliquée, ENSTA Paris, CNRS, Ecole Polytechnique, Institut Polytechnique de Paris, 91762 Palaiseau, France}%

\author{J. P. Couperus Cabada{\u{g}}}
\affiliation{Helmholtz-Zentrum Dresden--Rossendorf, Bautzner Landstrasse 400, 01328 Dresden, Germany}%

\author{A. Debus}
\affiliation{Helmholtz-Zentrum Dresden--Rossendorf, Bautzner Landstrasse 400, 01328 Dresden, Germany}%

\author{M. F. Gilljohann}
\affiliation{Ludwig--Maximilians--Universit{\"a}t M{\"u}nchen, Am Coulombwall 1, 85748 Garching, Germany}%
\affiliation{Laboratoire d’Optique Appliquée, ENSTA Paris, CNRS, Ecole Polytechnique, Institut Polytechnique de Paris, 91762 Palaiseau, France}%

\author{A. F. Habib}
\affiliation{University of Strathclyde, 107 Rottenrow, Glasgow G4 0NG, United Kingdom}

\author{T. Heinemann}
\affiliation{University of Strathclyde, 107 Rottenrow, Glasgow G4 0NG, United Kingdom}
\affiliation{The Cockcroft Institute, Keckwick Lane, Warrington WA4 4AD, United Kingdom}%

\author{B. Hidding}
\affiliation{The Cockcroft Institute, Keckwick Lane, Warrington WA4 4AD, United Kingdom}%
\affiliation{University of Strathclyde, 107 Rottenrow, Glasgow G4 0NG, United Kingdom}

\author{A. Irman}
\affiliation{Helmholtz-Zentrum Dresden--Rossendorf, Bautzner Landstrasse 400, 01328 Dresden, Germany}%

\author{F. Irshad}
\affiliation{Ludwig--Maximilians--Universit{\"a}t M{\"u}nchen, Am Coulombwall 1, 85748 Garching, Germany}%

\author{A. Knetsch}
\affiliation{Laboratoire d’Optique Appliquée, ENSTA Paris, CNRS, Ecole Polytechnique, Institut Polytechnique de Paris, 91762 Palaiseau, France}%

\author{O. Kononenko}
\affiliation{Laboratoire d’Optique Appliquée, ENSTA Paris, CNRS, Ecole Polytechnique, Institut Polytechnique de Paris, 91762 Palaiseau, France}%

\author{A. Martinez de la Ossa}
\affiliation{Deutsches Elektronen-Synchrotron DESY, Notkestraße 85, 22607 Hamburg, Germany}%

\author{A. Nutter}
\affiliation{Helmholtz-Zentrum Dresden--Rossendorf, Bautzner Landstrasse 400, 01328 Dresden, Germany}%
\affiliation{University of Strathclyde, 107 Rottenrow, Glasgow G4 0NG, United Kingdom}

\author{R. Pausch}
\affiliation{Helmholtz-Zentrum Dresden--Rossendorf, Bautzner Landstrasse 400, 01328 Dresden, Germany}%
\affiliation{Technische Universität Dresden, 01062 Dresden, Germany}

\author{G. Schilling}
\affiliation{Ludwig--Maximilians--Universit{\"a}t M{\"u}nchen, Am Coulombwall 1, 85748 Garching, Germany}%

\author{A. Schletter}
\affiliation{Ludwig--Maximilians--Universit{\"a}t M{\"u}nchen, Am Coulombwall 1, 85748 Garching, Germany}%
\affiliation{Technische Universit{\"a}t M{\"u}nchen, James-Franck-Str. 1, 85748 Garching, Germany}%

\author{S. Sch{\"o}bel}
\affiliation{Helmholtz-Zentrum Dresden--Rossendorf, Bautzner Landstrasse 400, 01328 Dresden, Germany}%
\affiliation{Technische Universität Dresden, 01062 Dresden, Germany}

\author{U. Schramm}
\affiliation{Helmholtz-Zentrum Dresden--Rossendorf, Bautzner Landstrasse 400, 01328 Dresden, Germany}
\affiliation{Technische Universität Dresden, 01062 Dresden, Germany}%

\author{E. Travac}
\affiliation{Ludwig--Maximilians--Universit{\"a}t M{\"u}nchen, Am Coulombwall 1, 85748 Garching, Germany}%

\author{P. Ufer}
\affiliation{Helmholtz-Zentrum Dresden--Rossendorf, Bautzner Landstrasse 400, 01328 Dresden, Germany}%
\affiliation{Technische Universität Dresden, 01062 Dresden, Germany}

\author{S. Karsch}
\email{stefan.karsch@physik.uni-muenchen.de}
\affiliation{Ludwig--Maximilians--Universit{\"a}t M{\"u}nchen, Am Coulombwall 1, 85748 Garching, Germany}%
\affiliation{Max Planck Institut für Quantenoptik, Hans-Kopfermann-Strasse 1, 85748 Garching , Germany}%


\maketitle

\onecolumngrid

\tableofcontents

\setcounter{equation}{0}
\setcounter{figure}{0}
\setcounter{table}{0}
\makeatletter
\renewcommand{\theequation}{S\arabic{equation}}
\renewcommand{\thefigure}{S\arabic{figure}}
\renewcommand{\thetable}{S\arabic{table}}

\section*{Supplemental Material}

\subsection{Details on the experimental setup}

\subsubsection{ATLAS-3000 laser system\label{laser}}
The experiment was performed at the ATLAS-3000 Ti:Sa chirped pulse amplification laser system at the Centre for Advanced Laser Applications (CALA) in Garching, Germany. The pulse length is measured to be $\SI{30} {\femto\second}$ (FWHM) at a central wavelength of $\SI{800} {\nano\meter}$. The pulse energy on target is on the order of $\SI{5} {\joule}$ for this experiment. The pulse is focused by a F/33 off-axis parabolic mirror. A nearly diffraction-limited focus is achieved by the use of a home-built Shack-Hartmann wavefront sensor in a closed loop with a commercial deformable mirror.

The energy on target has a large error margin of $(5\pm 1)\:\si{\joule}$. This is due to the fact that the transmission of compressor and beamline is affected by a rapid blackening of the optics. 
The exact transmission at the time of the different experiments was not recorded.

\subsubsection{Double-jet target}
The target setup for the staged hybrid LWFA-PWFA experiment consists of two Laval nozzles. Both nozzles are round and have exit diameters of $\SI{5}{\milli\meter}$ (LWFA) and $\SI{4}{\milli\meter}$ or $\SI{7}{\milli\meter}$ (PWFA), respectively. They create supersonic gas jets of adjustable gas density. The first nozzle is fed with a gas mixture of 96\% $\text{H}_2$ and 4\% $\text{N}_2$. A silicon wafer in the supersonic gas flow creates a sharp density down-ramp to utilize shock-front assisted ionisation injection in the LWFA stage. The LWFA stage is operated at a plateau plasma density of $1.3\times 10^{18}\:\si{\cm^{-3}}$. The stage separation can be varied between 0 and $\SI{5}{\centi\meter}$ and was, if not stated otherwise, $\SI{10}{\milli\meter}$ for the data presented in the article. This gives enough space for a removable laser blocker tape between both jets. The distance between the end of the first jet, the tape and the start of the second jet can be varied independently and was $\SI{7}{\milli\meter}$ and $\SI{3}{\milli\meter}$ respectively for the data shown in the article. The tape consists of $\SI{25} {\micro\meter}$ thick polyimide (Kapton\textsuperscript{\textregistered}) and is automatically advanced after every shot. To protect the laser from back reflection, the tape is tilted such that reflected light is not directed within the opening angle of the focusing parabola. Data shown in Fig.~\ref{stablerun} and Fig.~\ref{high_efficiency} was taken with laser blocker tape, while all other experiments were done without laser blocker. An auxiliary beam can be used to preionize the PWFA stage. For this purpose, $\SI{50} {mJ}$ picked from the main-beam are focused under a shallow angle into the PWFA stage using a $\SI{3}{\meter}$ (f/100) focal length spherical mirror. The relative arrival time of the electron driver and the preionizer laser can be varied and is set to values on the order of $\SI{10} {ps}$.

\begin{figure}[t]
  \centering
  \includegraphics*[width=0.7\linewidth]{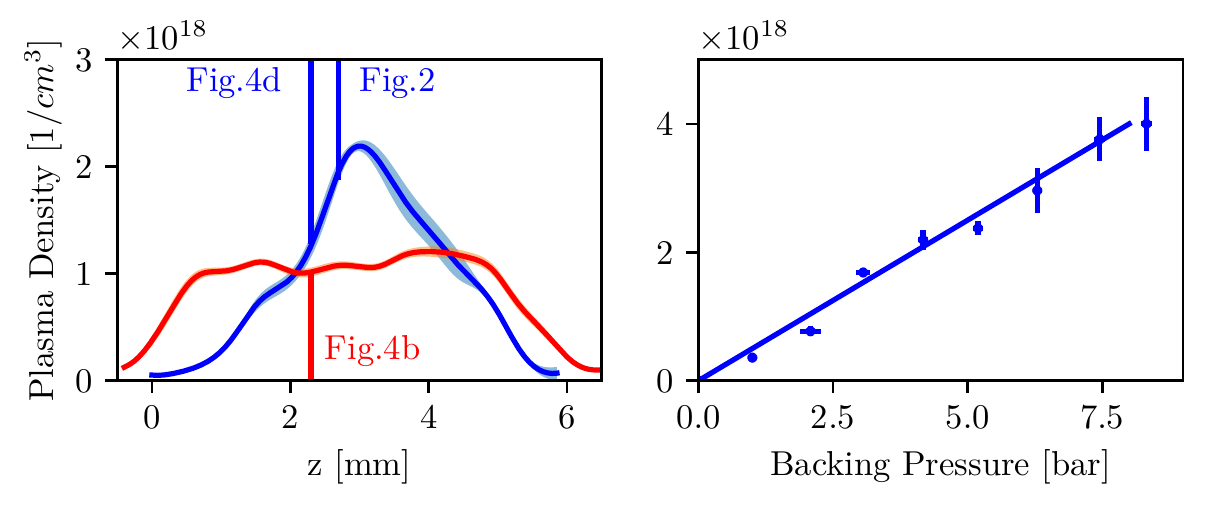}
\caption{\textbf{Plasma density of the PWFA stage}. Left: Lineout of the plasma density along the laser propagation axis for the $\SI{4}{mm}$-diameter (blue) and $\SI{7}{mm}$-diameter (red) nozzle used in the experiment. The density measurements are taken under experimental condition but with no density down-ramp for injection created. The position of the vertical lines corresponds to the approximate positions of the injection down-ramp for the different experiments presented in the manuscript. The height and downramp length of the density step is separately estimated from theoretical considerations. The density was measured using a Nomarski interferometer. Right: Scaling of plasma density in the $\SI{4}{mm}$-nozzle with backing pressure.} 
  \label{pressure_and_density}
\end{figure}

\subsubsection{Measured density of PWFA stage}

The density profile was retrieved using an Abel inversion of the phase information obtained with a Nomarski interferometer. The measured density was confirmed by plasma-wave imaging. 
Due to spatial constraints the electron propagation axis is aligned to be $\SI{5} {mm}$ from the nozzle exit. In particular for the $\SI{4}{mm}$-diameter nozzle, this leads to a bell-shaped density profile along the propagation axis. The vertical lines in Fig.~\ref{pressure_and_density} (left) correspond to the positions of the injection down-ramp for the different experiments presented in the main part of the manuscript. 

The injection position can only be varied within an interval of $\SI{500} {\micro\meter}$. Otherwise no injection is observed. The bell-shaped density profile might also explain the moderate energy gain since electrons injected at the density transition will soon begin to dephase in the down-ramp of the density profile. 

For the experiment demonstrating stable PWFA a 1:1 mixture of hydrogen and helium gas was used. Due to the use of the laser blocker tape for the main laser the gas in the second jet can not be assumed to be fully ionized. From a scan of the preionizer-beam intensity in pure helium it is confirmed that the first ionization level of helium is accessed at the intensity used during the experiment. At the time of arrival of the LWFA-generated drive beam we thus expect a mixture of fully ionized hydrogen and $\text{He}^+$ and therefore essentially the same plasma density vs. backing pressure scaling as in the pure hydrogen case.

\begin{figure}[t]
  \centering
  \includegraphics*[width=0.6\linewidth]{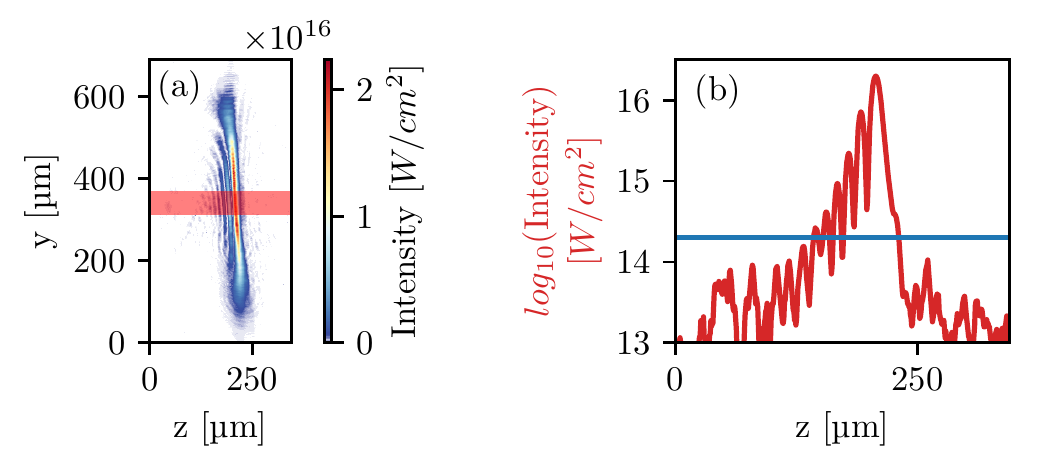}
  \caption{\textbf{Intensity distribution of the injector beam.} (a) Intensity distribution in the focal plane of the injector beam parallel to the wakefields propagation axis (red). (b) Intensity lineout along wakefield axis together with the assumed ionisation threshold (blue line).}
  \label{injector_beam}
\end{figure}

\begin{figure}[t]
  \centering
  \includegraphics*[width=\linewidth]{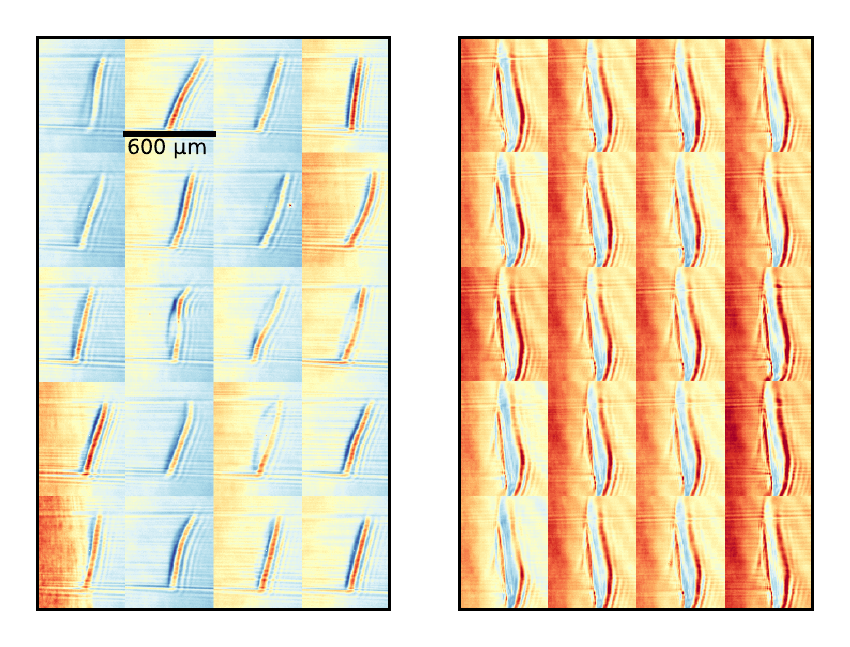}
  \caption{\textbf{Comparison of wire generated (left) and optically generated (right) shock for down-ramp injection.} For both cases 20 consecutive shots imaged with a transverse probe-beam via phase-contrast shadowgraphy are shown. Both position $\left(\SI{\pm 28} {\micro\meter}\right)$ and shape vary in the case of the wire generated shock. In particular the on-axis-angle between wakefield-driver and shock is subject to strong-shot-to-shot fluctuations. For the optically generated shock the position varies by less than the imaging resolution of $\SI{6}{\micro\meter}$. In this depiction the driver propagates from left to right. The axis of the wakefield is vertically centered in the images.}
  \label{shocks}
\end{figure}

\subsection{Down-ramp generation in PWFA stage}
\subsubsection{Comparing wire and optically generated hydrodynamic density down-ramps\label{supp_injection_methods}}

Two mechanisms to generate a hydrodynamic density down-ramp for injection in the PWFA stage have been investigated. They are compared in terms of the stability of their shape and position.

In preparatory studies for this work we generated a density down-ramp by placing a 200-$\si{\micro\meter}$-wide wire just inside the leading edge of the supersonic gas flow of the PWFA stage. The obstacle generates two hydrodynamic shocks, of which only one is situated in the density plateau of the jet.  
The wire is supported by two brackets separated by as far as $\SI{5}{\centi\meter}$. This mounting was necessary due to geometrical constraints. 
In this configuration we notice strong shot-to-shot fluctuations of charge and energy of an injected witness beam.  In Fig.~\ref{shocks} (left) the phase contrast shadowgraphs of the shock-front of 20 consecutive shots is presented. We observe fluctuations of the shape and the location of the shock (Fig.~\ref{shocks}, left) of $\pm \SI{28}{\micro\meter}$ (std), which most likely arise from mechanical vibrations of the wire in the supersonic gas flow. 

The alternative mechanism for the down-ramp generation, used in the main part of the this article, consists of an optical injector beam for locally ionizing and heating the plasma. The hot electron population expands and quickly diffuses. This leaves behind a positively charged ion-dominated region and thus, a radial electric field. Once both populations reach equilibrium, they propagate away from the heated region at the speed of sound $c_s=\sqrt{Z k_b T_e/m_i}$, which is of the order of $\sim 10\:\si{\micro\meter\per\nano\second}$ (for hydrogen at $T_e=\SI{20000}{K} \stackrel{\wedge}{\approx} \SI{2}{eV}$~\cite{Shalloo:2018fy}). This is higher than the speed of sound in the surrounding cold, neutral gas ($c_s = \sqrt{k_bT_0/m_i}$) and therefore leads to a shock-wave.

For this setup $\SI{50}{\milli\J}$ from the main laser-beam are picked and individually delivered to the target. This beam is perpendicularly focused onto the propagation axis of the electrons in the PWFA stage. The arrival time of the injector laser beam can be varied between 0 and $\SI{2} {ns}$ and was $\SI{1.3} {ns}$ in the experiments presented here. To guarantee perfect overlap of the injector and the electrons despite their non-negligible pointing jitter an astigmatic focusing geometry is chosen. The focusing optic is a  $\SI{15} {\centi\meter}$ focal length spherical mirror used under an angle of incidence of $15^\circ$. The resulting intensity distribution in the plane of the electron-beam is shown in Fig.~\ref{injector_beam}a. We deduce a peak intensity of $\SI{2e16}{\watt\per\cm^2}$ of the $(15 \times 550)\: \si{\micro\meter}^2$ (FWHM) focal spot. An intensity lineout along the axis of wakefield propagation (z) is given in Fig.~\ref{injector_beam}b. Assuming the Hydrogen to be ionized for a laser intensity  $I_{laser}>2\times10^{14}\frac{W}{cm^2}$ (blue line), the ionized region is expected to be roughly $\SI{100}{\micro\meter}$ wide. Thus after an expansion on a timescale of 1 ns the distance between both shocks will also be on the order of $\SI{100}{\micro\meter}$.

The increased stability of the optically induced hydrodynamic shock is presented in Fig.~\ref{shocks} (right). 
The position jitter of the optical injector beam is lower than the imaging resolution of $\SI{6}{\micro\meter}$. No shot-to-shot deformation of the shock is observed.

\subsection{PIC Simulations}
\subsubsection{Input parameters of PIC simulations}
The simulations use a lab-frame resolution of $\Delta z = \SI{50}{\nano\m}$, $\Delta r = \SI{200}{\nano\m}$, $m=3$ modes, a boosted frame \cite{Kirchen:2016iw} with $\gamma_{boost}=10$ and a particle density of $n_z\times n_r \times n_\theta = 10\times 4 \times 16$ within the injection region. The drive beam and plasma electrons making up the witness beam are initiated as two distinct species and can thus be separated in post-processing. As input we assume an electron-beam with a normally-distributed phase space, with a bunch length $c\sigma_t = \SI{3.5}{\micro\m}$, a radius $\sigma_r=\SI{5}{\micro\m}$ and a normalized emittance of $\epsilon_n=\SI{1}{mm}\:\si{\milli\radian}$. For our reference case we assume a central beam-energy of $E_0 = \SI{287}{\mega\eV}$, an energy spread $\sigma_E/E_0 = 10\%$ and a beam-charge of $Q_0 = \SI{340}{\pico\coulomb}$ modelling the driver from the first experiment. These parameters are varied according to the charge and energy fluctuations from the experiment by sampling a normal distribution centered around $Q_0$ and $E_0$, respectively.

The density profile used for the PIC simulations is shown in Fig. \ref{simulations}a of the main part. The density profile is a simplified version of the experimental measurement shown in Fig.~\ref{pressure_and_density}. A witness beam is injected in the final, $\SI{50}{\micro\m}$-long downramp of the density perturbation, placed on top of the density profile. The shape and size of the density perturbation was chosen according to the theoretical considerations explained above. 

Note that the parameters of the driver beam in our simulations are similar to our experiment and thus represent realistic experimental conditions. However, we do not explicitly model the full conditions underlying the experimental results presented in this article. The reasons for this are twofold. First, starting with the PWFA stage allows us to directly study the driver's influence on witness stability. Second, developing a 'digital twin' of the experiment would only be meaningful if it could produce additional insight, i.e. provide an accurate and quantitative representation of the experiment. This would require start-to-end simulations of the two stages, because the experiment only provides partial information about the 6-dimensional electron beam phase space, which is needed as initialization of the second stage. At this point such simulations are however not possible for us, because we cannot accurately model neither the blocking foil in FBPIC (because of the locally overdense plasma) nor the optically-generated density downramp. Accurate simulation of the latter would require a code that takes into account both tunneling ionization of molecular hydrogen and collision effects. We have therefore chosen a case of a more general density downramp, which will qualitatively be similar to both optically and hydrodynamically generated downramps and thus, is adequate for the subject of interest, namely the witness-stability variations with respect to the driver parameters.

\subsubsection{Driver emittance scan in PIC simulations}

\begin{figure*}[t]
  \centering
  \includegraphics*[width=\linewidth]{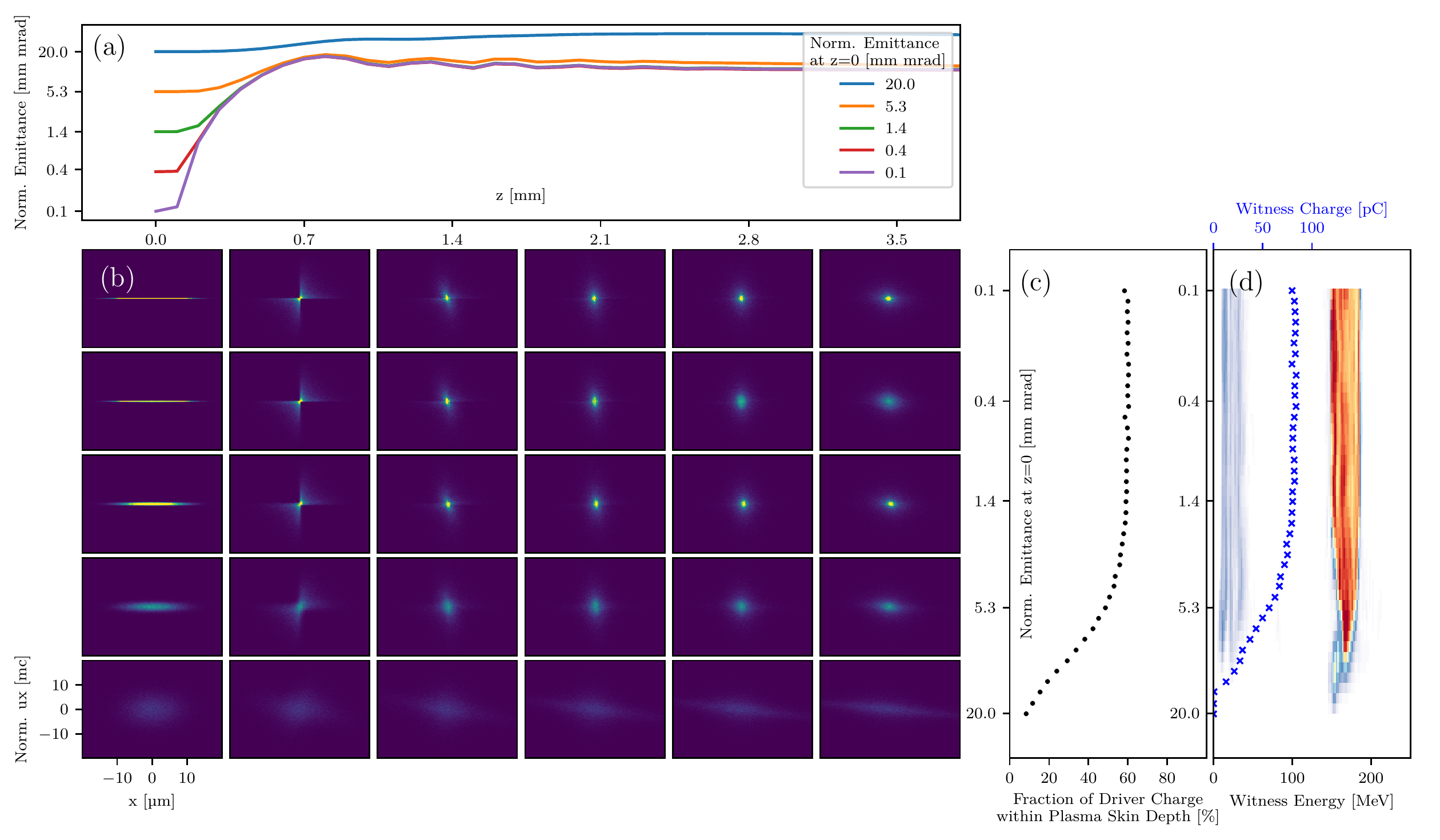}
\caption{\textbf{Simulation results on a scan of the driver emittance.} In (b) the x-ux phase space of the drive beam is plotted as a function of its initial normalized emittance and its propagation distance z in the PWFA. For all initial emittance values $<\SI{5}{mm}~\si{mrad}$ the driver evolution (a) is very similar and the emittance saturates at a level of $\sim\SI{10}{mm}~\si{mrad}$, given by the fact that the beam is not matched to the local betatron oscillation when entering in the plasma. Both the amount of driver charge refocused in the plasma to a spot smaller than the plasma skin depth (c) and the amount of injected witness charge (d) are mostly constant as long as the initial driver emittance is $<\SI{5}{mm}~\si{mrad}$. } 
  \label{Emittance_scan}
\end{figure*}

The driver emittance scan mimics the influence of the laser blocker tape on the driver electron bunch. For the simulations the drive beam radius was kept constant at $\sigma_r=\SI{5}{\micro\m}$ while the divergence and thus, the normalized emittance ($\epsilon_n=0.1-20~\si{mm}\:\si{\milli\radian}$) was varied. The size of the bunch is kept constant in the simulations because it is mainly given by the free space propagation of a few mm between LWFA and laser blocker. By just varying the divergence higher order distortions of the phase space due to the current filamentation are neglected that might dominate shortly after the interaction with the laser blocker tape. An estimate of the driver emittance in our experiment can be made assuming that the interaction at the laser blocker tape completely disturbs the phase-space correlation established during free space propagation. In this case the normalized emittance of the driver after the laser blocker tape can be approximated as $\epsilon_{\text{n,max}}=\gamma\sigma_{\theta, \text{ after tape}}\times \sigma_{x,\text{ at tape}}$ and is on the order of $500\times \SI{2}{mrad}\:\SI{5}{\micro \meter} = \SI{5}{\milli\meter\:\milli\radian}$. Here $\sigma_{x,\text{ at tape}}$ is estimated from the distance between LWFA stage and the laser blocker tape and the divergence of the drive beam from experiments without tape. $\sigma_{\theta, \text{ after tape}}$ is the typical driver divergence from reference shots with laser blocker tape but without PWFA stage.

As seen in Fig.~\ref{Emittance_scan}c, the amount of charged refocused to a diameter of less than one plasma skin depth is insensitive to the driver emittance in a range between 0.1 and 5 mm~mrad. For higher values of the driver emittance the amount of captured charge drops very quickly and thus, the wakefield strength drops very quickly.

The amount of injected charge has a very similar scaling as the amount of driver charge contributing to setting up the wakefield (Fig.~\ref{Emittance_scan}d). 

The drive beam is sent into the PWFA stage without using any refocusing device. Thus, after a free space drift of several mm, the phase space distribution of the witness is not matched to the local plasma conditions when entering the PWFA. Therefore the initial driver emittance is not conserved during the interaction with the plasma in the PWFA stage. At our experimental conditions it quickly degrades to a value around 10 mm mrad (see Fig.~\ref{Emittance_scan}a). However, this level seems to be only given by the betatron motion inside the plasma and independent of the initial driver emittance. Only for very large initial emittance of larger 10 mm.mrad the evolution in the plasma is drastically different and the driver degrades even further. The  evolution in x-ux phase space of the driver in the PWFA stage for different levels of initial emittance is visualized in Fig.~\ref{Emittance_scan}b. 

Our simulations show the insensitivity of the PWFA on the driver emittance over a large range of realistic emittance values of a LWFA-generated electron bunch. In particular this simulations can explain the equally high quality and similar energy gain of the witness beams in Fig.~\ref{angular_spectral_density}b+d of the main part.

The lower energy gain in experiments with laser blocker tape might be interpreted as a drop of wakefield strength as we observe in the simulations for $>\SI{5}{mm}~\si{mrad}$. However, due to the uncertainty about the exact driver phase space after the laser blocker and the exact density distribution of the PWFA we can't fully reproduce the low-energy witness spectra in our simulations. 

\subsubsection{Emittance of witness beams in PIC simulations}
\begin{figure*}[t]
  \centering
  \includegraphics*[width=0.6\linewidth]{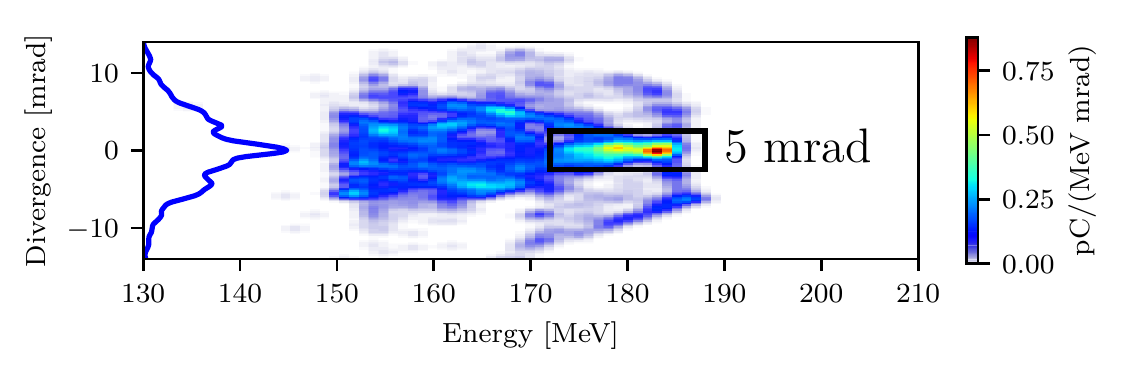}
\caption{\textbf{Simulated example of a witness spectrum as observed in a dipole spectrometer.} The beam  has a FWHM divergence of $\SI{1.5}{mrad}$. In addition to the narrow divergence feature there is a  high divergence background that is not observed in the experiment. Evaluating the normalized emittance in an opening angle of $\SI{5}{mrad}$ as indicated by the black box yields $\epsilon_{n,x} = \SI{0.28}{mm\times mrad}$ and $\epsilon_{n,y} = \SI{0.26}{mm\times mrad}$. } 
  \label{Emittance_witness}
\end{figure*}

A direct comparison of witness emittance in the experiment and in the PIC simulations is problematic. The transverse properties of the witness as predicted by the simulations do not fully match our experimental observations. In our experiments all witness charge is contained in one feature with a FWHM divergence of a few mrad (with laser blocker tape), or well below one mrad (without laser blocker tape). In simulations there is a high density feature with similarly small FWHM divergence of 1.5~mrad but there is also a much wider low charge density distribution with a divergence of more than 10~mrad (see Fig.~\ref{Emittance_witness} and Fig.~\ref{simulations}b in the main part).

If just picking the low divergence feature in the witness spectra for the analysis a normalized emittance of a few 0.1 mm mrad is obtained (e.g. 0.25 mm mrad if analyzing all electrons within an opening angle of 5 mrad, black box in Fig.~\ref{Emittance_witness}). Analyzing the full witness species this value increases to above 1 mm mrad. 

Note, that in the experiment the injector properties can be easily adjusted by change of the injector beam position, energy and timing. In PIC simulations on the other hand high-order parameter scans are computationally out of reach. Since our plasma diagnostic can only insufficiently characterize the density downramp height and shape, a generic plasma profile for injection based on theoretical considerations is simulated. 
We suspect that the transverse witness properties are not reproduced to its full extend because of the uncertainty about these input parameters. 

\subsubsection{Modelling loaded and unloaded PWFA in PIC simulations}

\begin{figure*}[t]
  \centering
  \includegraphics*[width=\linewidth]{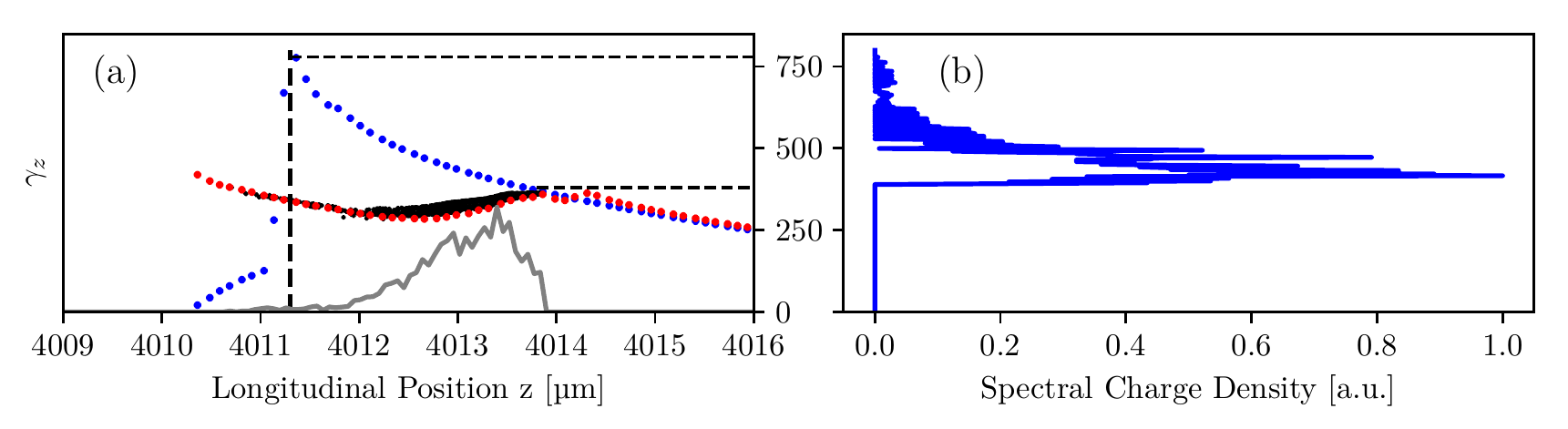}
\caption{\textbf{Hypothetical energy gain of the witness in an unloaded PWFA.} Dotted lines in (a) represent the energy gain of test particles as a function of its co-moving coordinate. The red dotted line corresponds to the simulation outcome with density downramp for witness injection. The phase space of the witness bunch (solid black) follows the red curve. The spectrum in (b) is calculated by projecting the witness current (gray in a.u.) on to the phase space of the test particles from the simulation without injection (blue dotted).} 
  \label{Unloaded_Wakefield}
\end{figure*}

Simulations have been performed to quantify the influence of beam loading on the energy gain of the witness in the PWFA stage. In addition to the driver and plasma electrons we initialize a further species of electrons in our PIC simulations. These test particles have a negligible charge density and are initialized with a high electron energy of $\SI{10}{GeV}$. They are arranged to serve as a probe for the longitudinal phase space in our simulations. To calculate the integrated energy gain as a function of the co-moving coordinate the difference in $\gamma_z$ of this particles in the last time step and the time step where witness injection occurs is calculated.
As seen in Fig.~\ref{Unloaded_Wakefield}a, for the simulations shown in the main part of this work this energy gain (red dotted line) is identical with the one of the injected witness (solid black).
The blue dotted line corresponds to test particles from an identical simulation, but without the density downramp for injection of a witness. A higher energy gain at positions behind the injected charge is clearly visible. An artificial spectrum for the unloaded case is calculated by projecting the witness current (gray line) from the simulations with injection onto the phase space of the test particles from the simulations without injection. Parts of the phase space that interacted with the 2nd bucket of the wakefield during parts of the simulation (corresponding to points beyond the kink at $\SI{4011.3}{\micro\meter}$ in Fig.~\ref{Unloaded_Wakefield}a) are discarded.

\subsection{High energy transfer efficiency}
\begin{figure}[t]
  \centering
  \includegraphics*[width=\linewidth]{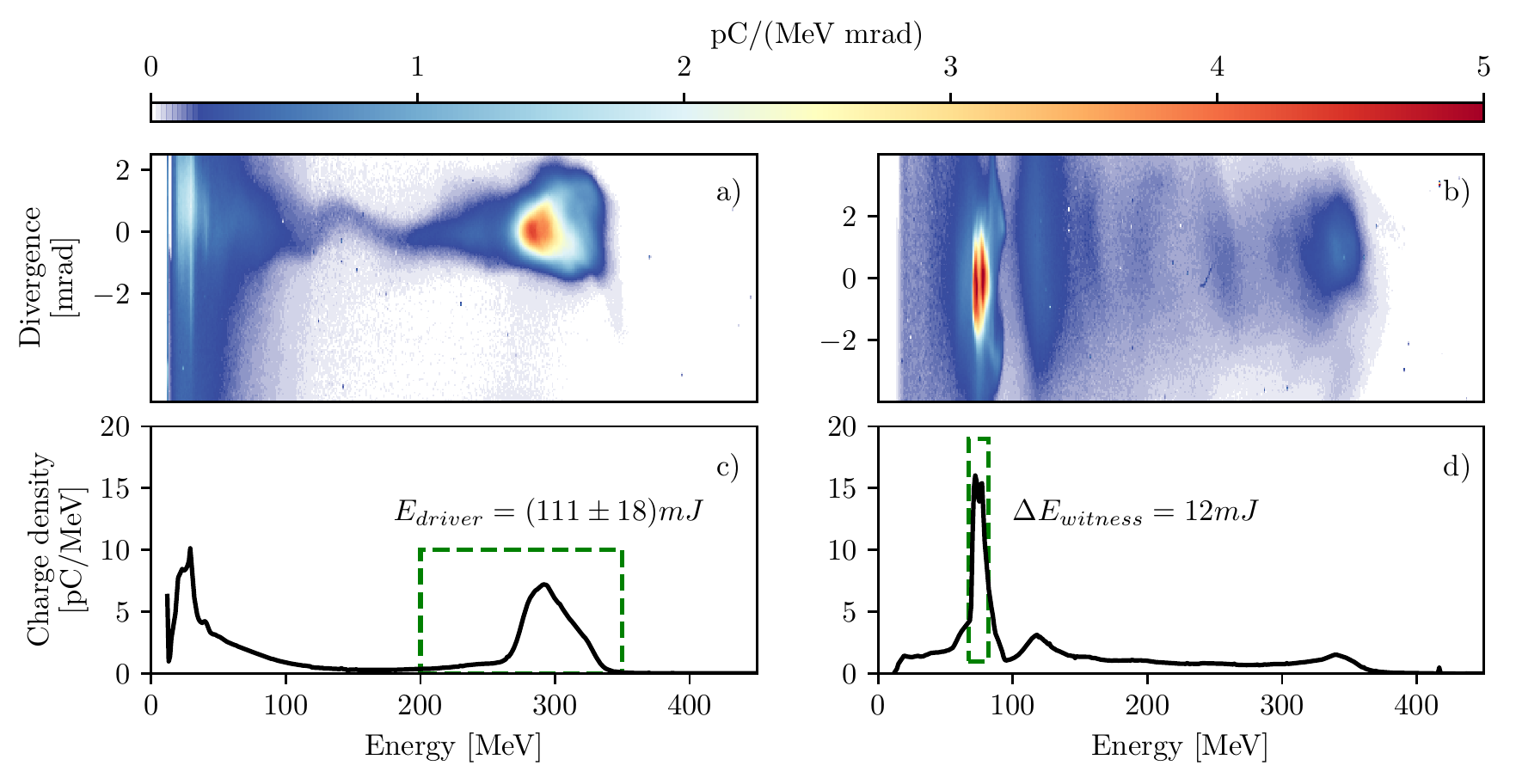}
  \caption{\textbf{Example spectra of a staged LWFA PWFA experiment with 10\% energy transfer efficiency.} (a) and (c) The LWFA bunches have a charge of $(300\pm100)\:\si{\pico\coulomb}$. (b) and (d) The charge of the shown witness bunch is $150\:\si{\pico\coulomb}$. The dashed lines indicate the spectral features used to calculate the energy transfer efficiency.}
  \label{high_efficiency}
\end{figure}

Figure \ref{high_efficiency} shows an example for a driver-witness-pair with an energy transfer efficiency as high as $10\%$. 

\subsection{Experiments without laser blocker}
\subsubsection{Identifying electron driver}
For the second set of experiments there is no laser blocker tape. Thus, there is an electron beam and an already diffracted laser in the second jet. We analyze their relative contribution to the wakefield in the second jet. With a measured vacuum divergence of $\SI{15}{\milli\radian}$ (half opening angle), the laser intensity is expected to drop by two orders of magnitude with respect to the first jet after $\SI{1}{\centi\meter}$ of propagation. This assumption neglects self-focusing in the first jet, which would cause an even smaller spot at the first plasma exit with consequently even larger divergence. A similar reduction in intensity has been observed for start-to-end PIC simulations \cite{Gotzfried:2020da}. Further assuming no laser-energy loss in the LWFA stage this leads to an upper boundary for the normalized vector potential of $a_0 \leq 0.34$ and hence a well non-relativistic beam. The electron-beam divergence is typically one order of magnitude smaller than the one of the laser. Furthermore, the electrons can self-focus at lower plasma densities than the laser leading to a high-density electron driver for the PWFA stage \cite{Gilljohann:2019kc}. 

\begin{figure}[t]
  \centering
  \includegraphics*[width=\linewidth]{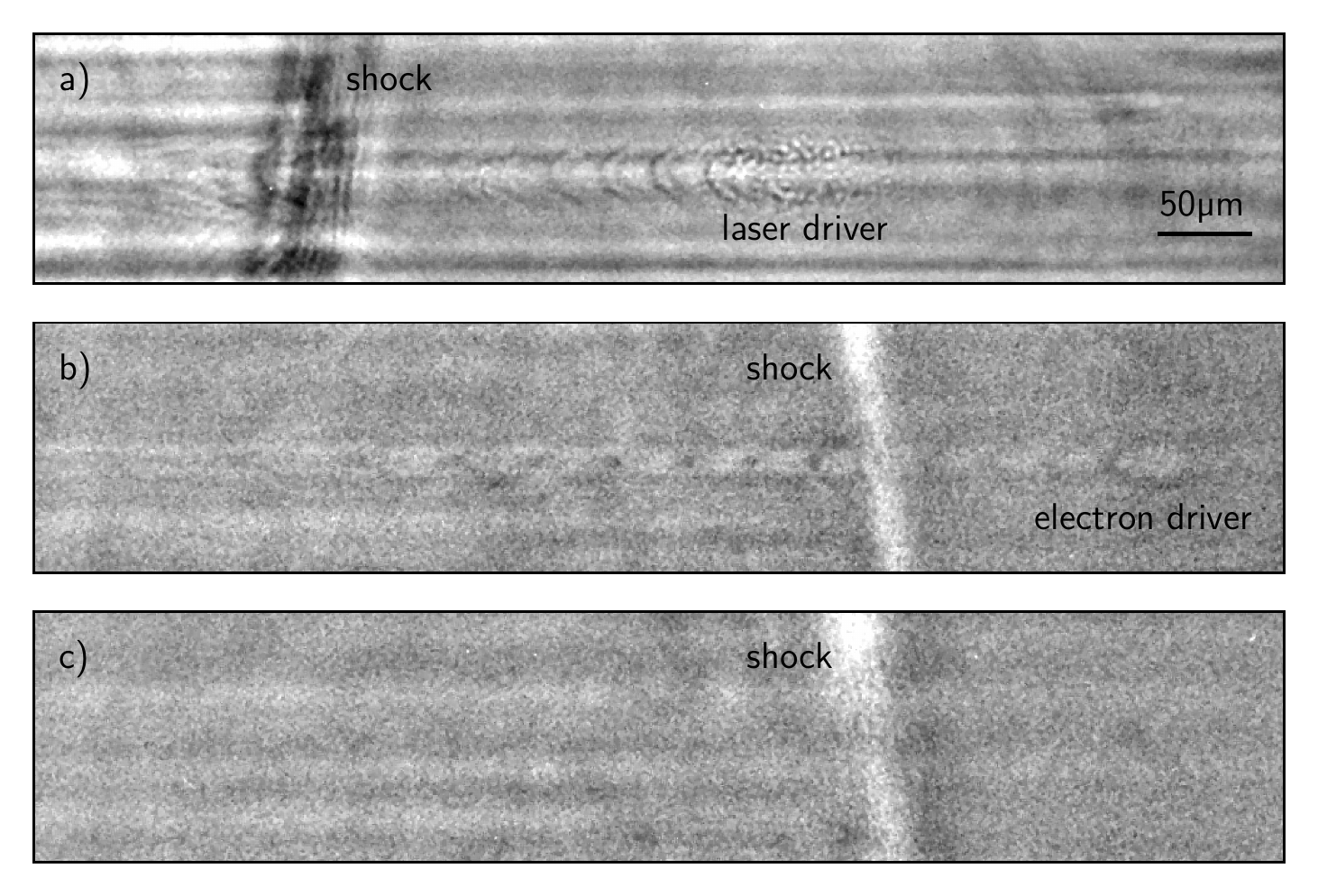}
  \caption{\textbf{Plasma-waves in the LWFA and the PWFA stage.} a: Laser driven plasma-wave in the LWFA stage after crossing through the shock-front. b: In the PWFA stage both the electrons from the LWFA stage and the diffracted and depleted laser are present. The wave train exhibits a distinct bubble-like feature at the position of the drive beam. c: The current of the driver in the PWFA stage is reduced by removing the blade for shock-front injection in the LWFA stage. Under else identical conditions no plasma wave is observed in this setup. Thus, we conclude the wakefield in the second stage is predominantly driven by the electrons. In all images the respective driver propagates from left to right. }
  \label{plasma_wave}
\end{figure}

To identify the LWFA-generated electron beam as the dominant driver of the wakefield in the PWFA stage plasma-wave images are analysed. Fig.~\ref{plasma_wave}a and b show the few-cycle shadowgraph of a plasma-wave in the first and in the second jet. In both cases the driver propagates from left to right. Also in both cases the respective density down-ramp for injection is visible. In the laser driven wakefield in the first jet a grainy feature leads the wave train (Fig.~\ref{plasma_wave}a), a typical observation for the case of a highly relativistic laser driver. Also the wakefield is flanked by off-axis striations extending further downstream than the peak of the laser driver. This is attributed to the laser being intense enough to ionise plasma already early in the leading edge and in off-axis intensity maxima. In the second jet a distinct bubble like feature leads the wave train (Fig.~\ref{plasma_wave}b), while a grainy feature as above is not present.  
In contrast, when the LWFA-generated drive bunch is switched off by removing the blade for shock-front injection in the LWFA stage at otherwise identical conditions, the plasma-wave in the second jet vanishes (Fig.~\ref{plasma_wave}c), indicating no observable laser-driven wakefield.

\subsubsection{Spectral features of driver and witness beam}
\begin{figure*}[t]
  \centering
  \includegraphics*[width=\linewidth]{ 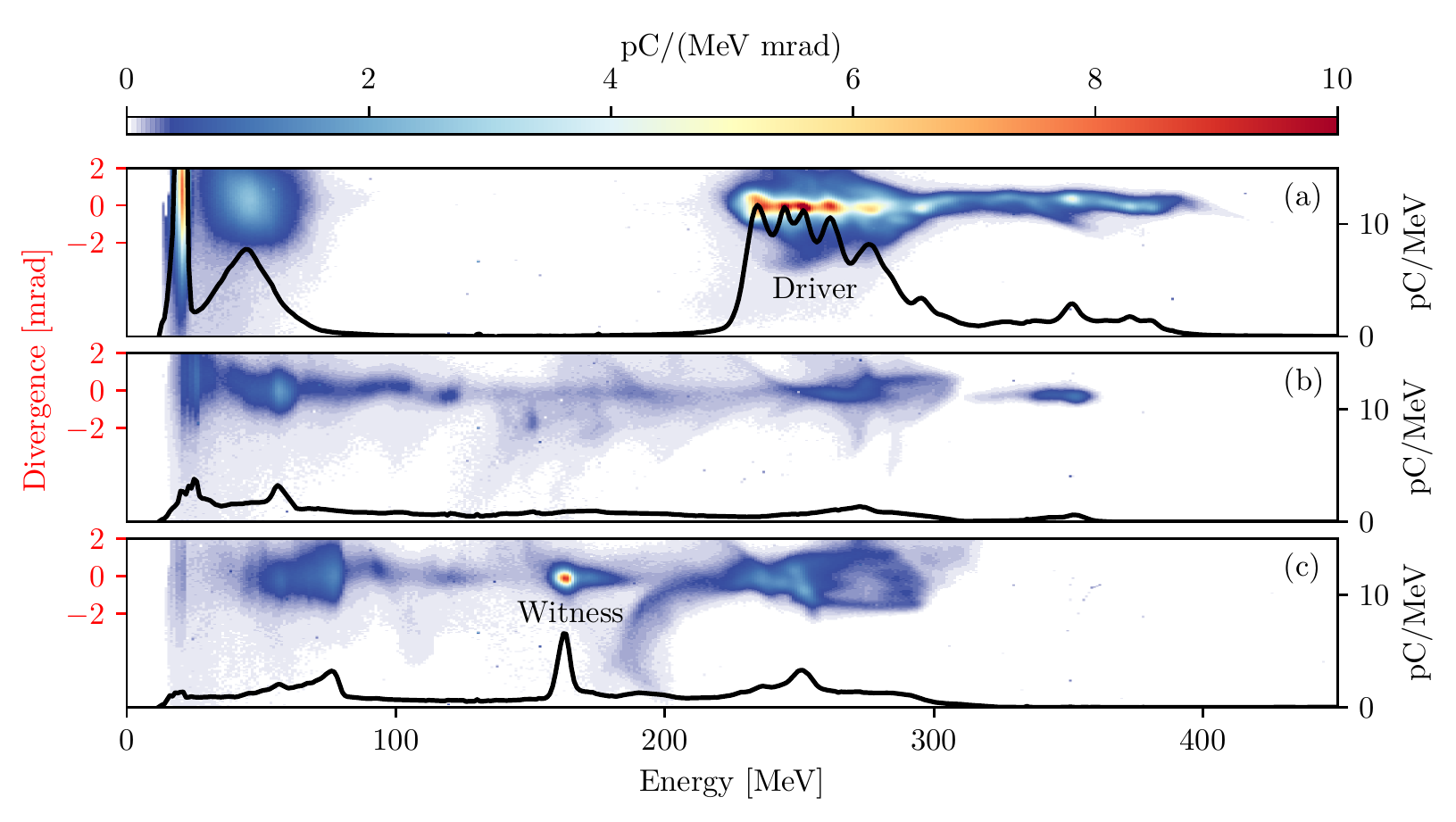}
\caption{\textbf{Spectral Features of LWFA and PWFA electron-bunches}. (a) Typical spectrum of a drive beam from Fig.~\ref{angular_spectral_density} of the main part. The low-energy feature at $\SI{50}{\mega eV}$ is attributed to a second injection event. (b) Spectrum after the PWFA stage without down-ramp for injection. In addition to the driver features there is a broadband background due to decelerated driver charge. (c) Spectrum after the PWFA stage with injector (Data from Fig.~\ref{angular_spectral_density} of the main part.). The witness bunch injected at the density down-ramp is visible at $\SI{162}{MeV}$ on top of the spectral features from (b).  }
  \label{spectral_features}
\end{figure*}

The spectra of our LWFA electron-bunches typically have additional low energy features. We note that spectral components below $\SI{100}{MeV}$ in Fig.~\ref{spectral_features}a are most likely the result of a secondary injection event in the LWFA stage. This component is usually not visible in experiments with laser blocker (cf. first section of main part) because low-energy electrons are strongly scattered at the tape and thus produce only a weak signal on the spectrometer. When switching on the second gas jet without shock for down-ramp creation there is a continuous electron background due to decelerated charge of the driver, see Fig.~\ref{spectral_features}b.

For an injector laser beam with a peak intensity of $\SI{2e16}{\watt\per\cm^2}$ and $\SI{10}{\milli\meter}$ gap between the jets, broad-band witness bunches are produced on every shot (representative: Fig.~\ref{angular_spectral_density_supp}d). To unambiguously identify the witness bunch and its origin the injector-beam is switched off and the distinct and dense feature on top of the broad-band background vanishes (as in Fig.~\ref{spectral_features}b). We thus conclude that the additional electron bunch visible after the PWFA stage is injected at the optically injected down-ramp in the PWFA stage. 
The charge of these broadband witness bunches is $(120\pm50)\:\si{\pico\coulomb}$. The large error margin is mainly given by the uncertainty about the decelerated driver charge because the witness spectrally overlaps with the decelerated part of the driver.
As seen in Fig.~\ref{angular_spectral_density_supp}d+e these witness bunches feature sub-millirad divergence over their whole spectrum and have a distinct high-energy cut-off energy at $(178\pm9)\:\si{MeV}$ (5\%, std in a set of 30 consecutive shots). 

\subsubsection{Increasing angular-spectral charge density}
\begin{figure*}[t]
  \centering
  \includegraphics*[width=\linewidth]{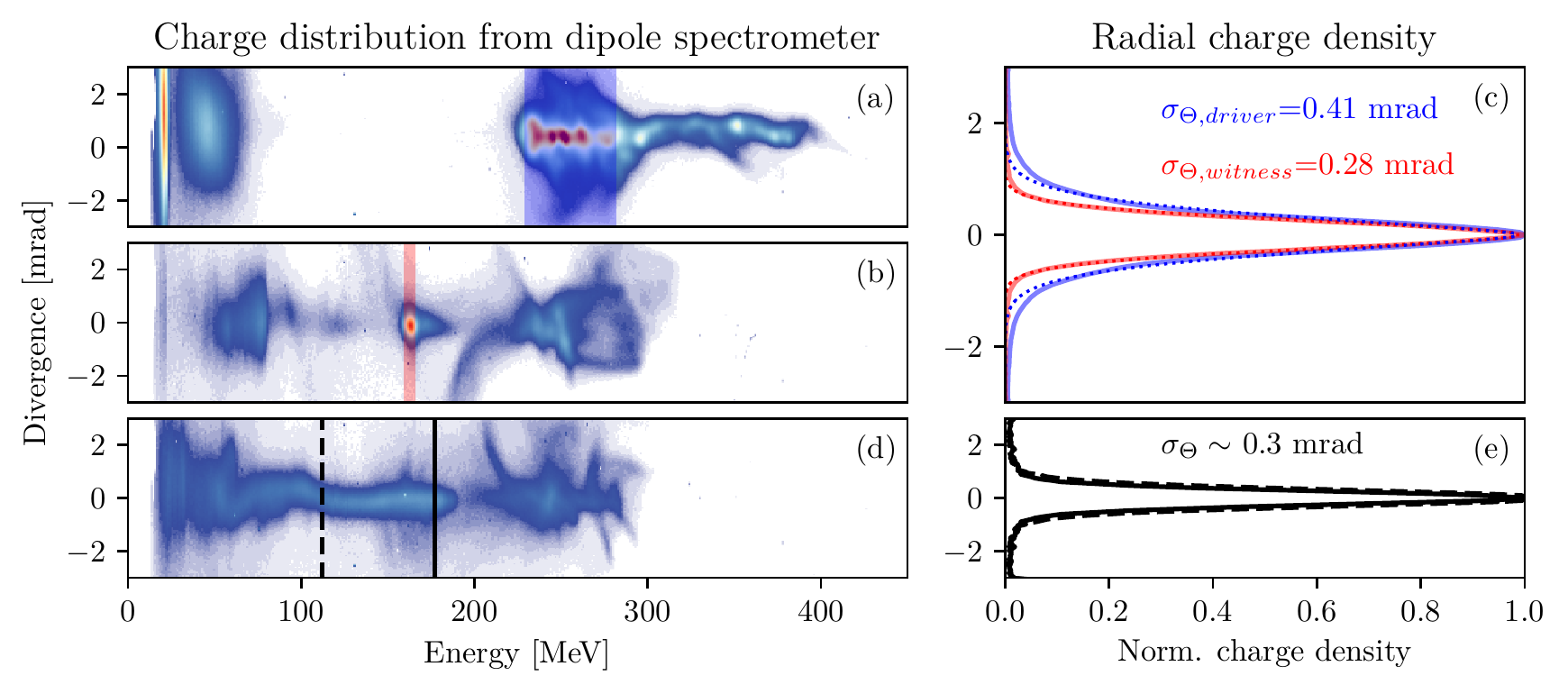}
\caption{\textbf{Angular-spectral charge density in the PWFA stage}. (a) Typical spectrum of a low-divergence drive beam from LWFA. (b) Spectrum after the PWFA stage with charge of the witness optimized for high charge density. (c) Divergence deduced from the radial charge distribution of the beams from a and b evaluated in the shaded energy intervals (FWHM) from a and b. While the drive beam exhibits a narrow-width core and high-divergence wings the charge of the witness beam is almost fully concentrated in one low-divergence feature. The thin dotted lines are super-Gaussian fits to deduce the RMS-divergence. (d) Broad-band spectrum after the PWFA stage with injection at the optically generated density down-ramp. (e) Divergence of the broad-band witness at two energy slices (overlapping). No modulation of the divergence along the energy spectrum is observed.}
  \label{angular_spectral_density_supp}
\end{figure*}

The energy spread of the witness bunches can be controlled by adjusting the injected charge. For a certain amount of injected charge ($\approx\SI{30}{pC}$) witness bunches with few \% energy spread are generated (Fig.\ref{angular_spectral_density_supp}b). At the same time their spectral charge density is 3 times higher than in the case of the higher charge but broadband beams (Fig.\ref{angular_spectral_density_supp}d). 

As a measure for the quality of the electron beams we introduce the angular-spectral charge density. We define it as the average charge density per energy interval within its FWHM energy spread and spatially within the RMS solid angle taken by the beam. 
Spectra obtained with a dipole spectrometer (as in Fig.~\ref{angular_spectral_density_supp}a+b+d) are spectrally resolved projections of the electron distribution along one spatial transverse dimension. We calculate the radial charge distribution of the electron bunches using an inverse Abel transformation (Fig.~\ref{angular_spectral_density_supp}c+e) and by that assuming radial symmetry of the bunches. It is a typical feature of our high angular-spectral charge density PWFA-beams that all charge is contained within a single dense feature with a divergence of $\sim \SI{0.5}{\milli\radian}$ (FWHM). Some LWFA-generated drive beams may exhibit a similar FWHM divergence. However, this is due to dense and little divergent sub-features on top of a lower density and more divergent (few mrad FWHM) charge distribution (compare Fig.~\ref{angular_spectral_density_supp}a-c). To better reflect these different transverse shapes of LWFA and PWFA beams, we deduce the divergence of the beams from the standard deviation of a super-Gaussian fit to the radial charge distribution (Fitting function: $f(x)\sim \text{exp}\left(\left[-x^2/(2\sigma^2) \right]^p\right)$, with $p<1$). 

\subsubsection{Emittance of witness beams injected via density down-ramp injection}

\begin{figure*}[t]
  \centering
  \includegraphics*[width=\linewidth]{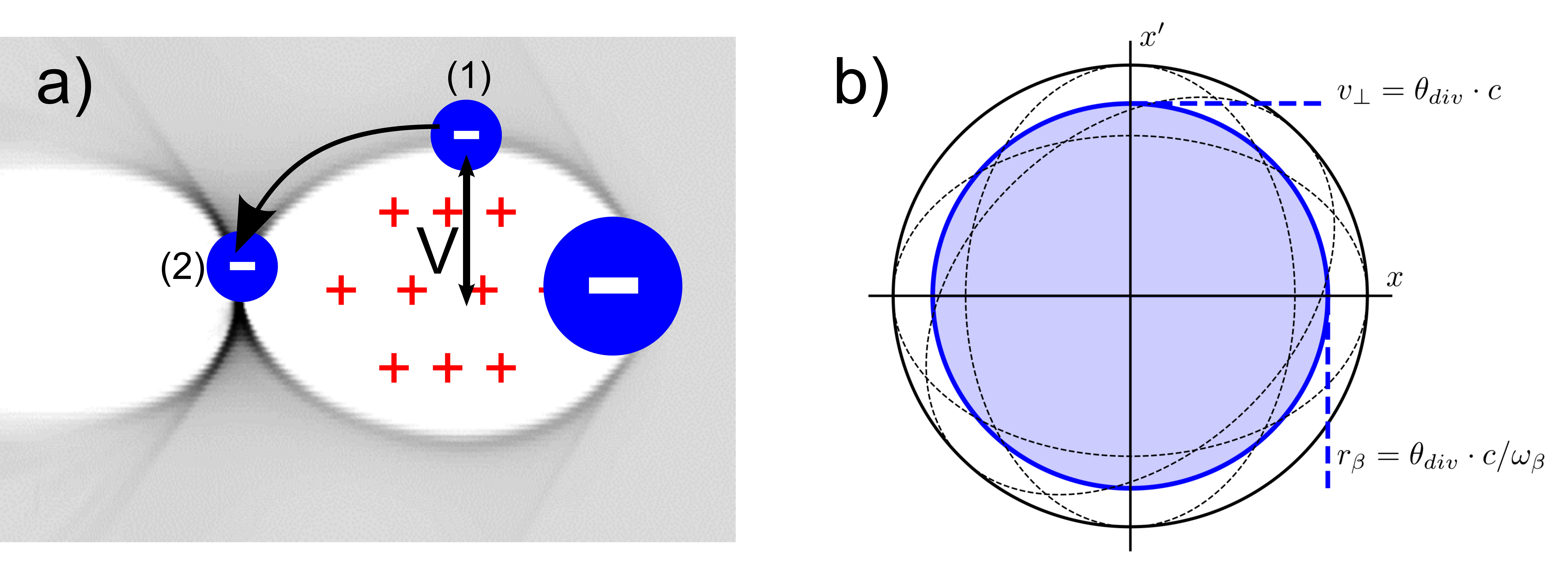}
\caption{\textbf{Transverse momentum of trapped electrons and their betatron trajectories}. a: Electrons forming the sheath of the \textit{bubble} (1) have been displaced against the Coulomb force of the ion-background. Falling back on axis (2) this potential energy is transferred into kinetic energy and thus a transverse momentum. b: The blue-shaded area is the (matched) distribution in transverse trace-space of an exemplary electron bunch. The solid blue line indicates the trace-space orbit of a single particle due to its betatron oscillation. The magnitude of both semi-axes is given by the particles initial transverse momentum. The black dotted lines are an exemplary unmatched trace-space distribution of same emittance plotted at different betatron phases.}
  \label{betatron}
\end{figure*}

We first estimate the order of magnitude of the transverse momentum of injected electrons. From the transverse momentum an upper bound for the normalized emittance is calculated. According to our simplified model the potential energy V of sheath electrons (Fig.~\ref{betatron}a (1)) in the field of the ion-bubble at maximum transverse displacement $r_b$ (Eq.\ref{blowout_radius})  is $V =  \int_{0}^{r_b}F_{\text{ion}}(r) dr = e^2 n_0/4\epsilon_0\int_{0}^{r_b}r dr  = \SI{30}{eV}\times I_0/\si{A}$. 

To estimate an upper boundary for the transverse momentum we neglect that plasma electrons forming the sheath also acquire substantial velocities in longitudinal direction. In a full model a electrostatic description would not be accurate. 
In particular, transverse momentum can be transferred to longitudinal one via the B-field. 
Assuming the potential energy is fully transferred into kinetic energy by falling back on axis (Fig.~\ref{betatron}a (2)), we calculate the corresponding transverse momentum and beam divergence. 
From the sum of energies $V+m_0c^2=\sqrt{c^2p_\perp^2+(m_0c^2)^2}$ a transverse momentum of $p_\perp=mc\sqrt{(V/mc^2+1)^2-1}$  is calculated. Electrons carrying the full transverse momentum can be assumed to be the most divergent ones after exiting the plasma. Assuming a uniform continuous distribution of the transverse momenta among the bunch-electrons the RMS-momentum of $p_{\perp,RMS} = p_\perp/\sqrt{3}$ can be calculated. 
Thus, at the end of the longitudinal acceleration the RMS divergence angle is $\sigma_\theta=p_{\perp,\text{RMS}}/p_{\parallel} = p_{\perp,\text{RMS}}/(300mc) = 4\times10^{-3}$, where we set $\gamma=300$ as in the experiment presented in Fig.~\ref{angular_spectral_density}b and a driver current of $I_0=\SI{20}{\kilo\ampere}$. 
Inside the bubble the transverse motion of an electron of the witness bunch is a harmonic oscillation at the betatron-frequency $\omega_\beta=\omega_{p}/\sqrt{2\gamma}$. The trace-space orbit of a trapped electron is schematically shown in Fig.~\ref{betatron}b (solid blue line). The trajectory of a given electron is determined by its transverse momentum. According to the theory of a harmonic oscillator the amplitude of the betatron oscillation is $r_{\text{betatron,RMS}}=  \sigma_{\theta}c\sqrt{2\gamma }/\omega_p$. For $\gamma = 300$ and $n_0=1\times10^{18}\:\si{cm^{-3}}$ the RMS betatron radius is $\SI{500}{nm}$. Electrons on the RMS betatron trajectory are moving on the phase space orbit that defines the contour of the phase space ellipse (Fig.~\ref{betatron}b blue shaded area). Thus, we calculate an upper limit for the normalized emittance of $\epsilon_{\text{n,max}}=\gamma\sigma_\theta\sigma_x=300\times \SI{4}{mrad}\:\SI{0.5}{\micro \meter} = \SI{0.6}{\milli\meter\:\milli\radian}$. 

To preserve the emittance of an electron-bunch both during its acceleration and extraction from the plasma its trace-space distribution should be matched to the local betatron oscillation \cite{Mehrling:2012ky}. 
The witness spectra with broad energy spread (Fig.~\ref{angular_spectral_density_supp}d+e) can be understood as a probe for the longitudinal phase-space. Other than in previous experiments \cite{CouperusCabadag:2021gj} there is no modulation of the beam divergence along the spectrum observed. 
This observation hints to a witness beam-size well matched to the local betatron-frequency throughout the acceleration and in particular during extraction from the plasma. Otherwise a situation as in Fig.~\ref{betatron}b (black dotted lines) is expected. Here the envelope of the unmatched trace-space distribution rotates while the betatron phase advances. This rotation depends on the local betatron frequency, i.e. on the electron energy and the local plasma frequency. Thus, for different longitudinal slices a different betatron phase advance is expected. This in turn would be visible as a modulation of the divergence along the spectrum of the electron-bunch.

\FloatBarrier

\bibliography{references.bib}